\documentclass{jpp}
\usepackage{graphicx}
\usepackage{epstopdf, epsfig}
\usepackage{amsmath}
\usepackage{mathtools}
\usepackage{color}
\usepackage[breaklinks,colorlinks=true,citecolor=cobalt,linkcolor=red,urlcolor=blue]{hyperref}
\usepackage[hyphenbreaks]{breakurl}

\definecolor{cobalt}{rgb}{0.0, 0.28, 0.67}
\definecolor{airforceblue}{rgb}{0.36, 0.54, 0.66}
\definecolor{armygreen}{rgb}{0.29, 0.33, 0.13}

\newcommand{\physrep}{Phys.~Rep.} 
\newcommand{\apj}{ApJ} 
\newcommand{\apjl}{ApJ} 
\newcommand{\apjs}{ApJ} 
\newcommand{\aj}{AJ} 
\newcommand{\aap}{A\&A} 
\newcommand{\araa}{ARAA} 
\newcommand{\mnras}{MNRAS} 
\newcommand{\nat}{Nature} 
\newcommand\aapr{A\&A~Rev.} 
\newcommand\ssr{SSRv} 
\newcommand\pre{PRE} 

\newcommand{\mach}{\mathcal{M}}

\newcommand{\machazero}{\mathcal{M}_\mathrm{A,0}}
\newcommand{\Bzero}{B_\mathrm{0}}
\newcommand{\Bturb}{B_\mathrm{turb}}

\newcommand{\eps}{\epsilon}

\newcommand{\ddt}[1]{\frac{\partial{#1}}{\partial t}}

\newcommand{\bfu}{\mathbf{u}}
\newcommand{\bfB}{\mathbf{B}}
\newcommand{\bfomega}{\boldsymbol{\omega}}
\newcommand{\vect}[1]{{\mathbf{#1}}}
\newcommand{\cs}{c_\mathrm{s}}

\newcommand{\Gauss}{\mathrm{G}}
\newcommand{\cm}{\mbox{cm}}
\newcommand{\g}{\mbox{g}}
\newcommand{\G}{\mbox{G}}
\newcommand{\ted}{t_\mathrm{ed}}
\newcommand{\Em}{E_{\mathrm{m}}}
\newcommand{\Emzero}{E_{\mathrm{m0}}}
\newcommand{\Ek}{E_{\mathrm{k}}}
\newcommand{\satlev}{\left(\Em/\Ek\right)_\mathrm{sat}}
\newcommand{\solratio}{E_\mathrm{sol}/E_\mathrm{tot}}
\newcommand{\Rm}{\mathrm{Rm}}
\newcommand{\Pm}{\mathrm{Pm}}

\newcommand{\re}{\mathrm{Re}}
\newcommand{\rmag}{\mathrm{Rm}}
\newcommand{\pmag}{\mathrm{Pm}}
\newcommand{\emag}{E_\mathrm{m}}
\newcommand{\ekin}{E_\mathrm{k}}
\newcommand{\esat}{(\emag/\ekin)_\mathrm{sat}}

\shorttitle{Magnetic fields in astrophysical plasmas}
\shortauthor{Christoph Federrath}

\title{Magnetic field amplification in turbulent astrophysical plasmas}

\author{Christoph Federrath\aff{1}
  \corresp{\email{\href{mailto:christoph.federrath@anu.edu.au}{christoph.federrath@anu.edu.au}}}}

\affiliation{\aff{1}Research School of Astronomy and Astrophysics, Australian National University, Canberra, ACT~2611, Australia}

\begin{document}

\maketitle

\begin{abstract}
Magnetic fields play an important role in astrophysical accretion discs, and in the interstellar and intergalactic medium. They drive jets, suppress fragmentation in star-forming clouds and can have a significant impact on the accretion rate of stars. However, the exact amplification mechanisms of cosmic magnetic fields remain relatively poorly understood. Here I start by reviewing recent advances in the numerical and theoretical modelling of the \emph{turbulent dynamo}, which may explain the origin of galactic and inter-galactic magnetic fields. While dynamo action was previously investigated in great detail for incompressible plasmas, I here place particular emphasis on highly compressible astrophysical plasmas, which are characterised by strong density fluctuations and shocks, such as the interstellar medium. I find that dynamo action works not only in subsonic plasmas, but also in highly supersonic, compressible plasmas, as well as for low and high magnetic Prandtl numbers. I further present new numerical simulations from which I determine the growth of the turbulent (un-ordered) magnetic field component ($\Bturb$) in the presence of weak and strong guide fields ($B_0$). I vary $B_0$ over 5 orders of magnitude and find that the dependence of $\Bturb$ on $B_0$ is relatively weak, and can be explained with a simple theoretical model in which the turbulence provides the energy to amplify $\Bturb$. Finally, I discuss some important implications of magnetic fields for the structure of accretion discs, the launching of jets, and the star formation rate of interstellar clouds.
\end{abstract}

\section{Introduction}

Magnetic fields are ubiquitous in the Universe. Examples include astrophysical accretion discs around young stars and active galactic nuclei, the interstellar and intergalactic medium, and even the early Universe when the first stars and galaxies formed.

The \emph{turbulent dynamo} mechanism is believed to be the main cause of cosmic magnetism \citep{BrandenburgSubramanian2005}. Magnetic fields are amplified exponentially via the turbulent dynamo, leading to dynamically important magnetic forces and energies on relatively short time scales. The amplification of the magnetic field arises from sequences of ``stretching, twisting, and folding'' of the field, until the magnetic field lines are so tightly packed that the magnetic energy density becomes comparable to the kinetic energy density provided by turbulent motions.

Dynamo action ranges from the Earth and the Sun \citep{CattaneoHughes2001}, over the interstellar medium to whole galaxies \citep{BeckEtAl1996,Beck2016}. The turbulent dynamo is important for the formation of the large-scale structure of the Universe \citep{RyuEtAl2008,MiniatiBell2011,IapichinoBrueggen2012,VazzaEtAl2014,MiniatiBeresnyak2015,BeresnyakMiniati2016}, in clusters of galaxies \citep{SubramanianShukurovHaugen2006} and in the formation of the first cosmological objects in dark matter haloes \citep{SchleicherEtAl2010}. It determines the growth of magnetic energy in solar convection \citep{CattaneoHughes2001,MollEtAl2001,GrahamEtAl2010}, in the interior of planets \citep{RobertsGlatzmaier2000} and is relevant for liquid metal experiments on Earth \citep{MonchauxEtAl2007}. It may further explain the far-infrared--radio correlation in spiral galaxies \citep{SchleicherBeck2013}. After the turbulent dynamo has amplified tiny seeds of the magnetic field, which can be generated during inflation, the electroweak or the QCD phase transition \citep{GrassoRubinstein2001}, the large-scale dynamo kicks in and generates the large-scale magnetic fields observed in planets, stars and galaxies today \citep{BeckEtAl1996,BrandenburgSubramanian2005}.

One of the most important distinctions that we have to make when considering turbulent gases, fluids and plasmas, is the level of \emph{compressibility} of the medium. For instance, in the Earth and the Sun, the dynamo is driven by subsonic, nearly incompressible flows. By contrast, interstellar clouds and galaxies are dominated by highly compressible turbulence \citep{MacLowKlessen2004,ElmegreenScalo2004,McKeeOstriker2007,HennebelleFalgarone2012,PadoanEtAl2014}. Indeed, the gas densities in the interstellar medium range from $\lesssim1$ particle per $\cm^3$ \citep[the average gas density in a Milky-Way type galaxy; see][]{Ferriere2001} to $\gtrsim10^{10}\,\cm^{-3}$ (where the gas becomes optically thick and proceeds to the formation of a star).

Numerical studies of non-driven turbulence demonstrated that supersonic turbulence decays quickly, in about a crossing time \citep{ScaloPumphrey1982,MacLowEtAl1998,StoneOstrikerGammie1998,MacLow1999}. Since we observe highly compressible interstellar turbulence, that turbulence must be driven by some physical forcing mechanisms. Those mechanisms include supernova explosions, ionizing shells from high-mass stellar feedback \citep{McKee1989,KrumholzMatznerMcKee2006,BalsaraEtAl2004,AvillezBreitschwerdt2005,BreitschwerdtEtAl2009,GritschnederEtAl2009,PetersEtAl2010,PetersEtAl2011,GoldbaumEtAl2011,LeeMurrayRahman2012}, gravitational collapse and accretion of material \citep{Hoyle1953,VazquezCantoLizano1998,KlessenHennebelle2010,ElmegreenBurkert2010,VazquezSemadeniEtAl2010,FederrathSurSchleicherBanerjeeKlessen2011,RobertsonGoldreich2012}, and galactic spiral-arm compression and cloud-cloud collisions \citep{DobbsBonnell2008,DobbsEtAl2008,TaskerTan2009,BenincasaEtAl2013}, as well as magneto-rotational instability \citep{PiontekOstriker2004,PiontekOstriker2007,TamburroEtAl2009}. Jets and outflows from young stellar objects \citep{NormanSilk1980,BanerjeeKlessenFendt2007,NakamuraLi2008,CunninghamEtAl2009,CarrollFrankBlackman2010,WangEtAl2010} and active galactic nuclei \citep{MukherjeeEtAl2016} also drive turbulence. Turbulence in high-redshift galaxies and their low-redshift analogues may be driven by feedback \citep{GreenEtAl2010,FisherEtAl2014}. A discussion and comparison of turbulence driving mechanisms is published in \citet{MacLowKlessen2004}, \citet{Elmegreen2009}, and \citet{FederrathEtAl2016iaus}.

Most importantly, the majority of turbulence drivers (e.g., supernova explosions, high-mass stellar winds, and accretion) are expected to drive compressible modes, so we refer to these as ``compressive drivers''. On the other hand, solenoidal modes can be generated directly by shear \citep{FederrathEtAl2016} and the MRI (so we call them ``solenoidal drivers''), and indirectly by nonlinear interactions of multiple colliding shock fronts \citep{Vishniac1994,SunTakayama2003,KritsukEtAl2007,FederrathDuvalKlessenSchmidtMacLow2010}, baroclinity \citep{PadoanEtAl2016}, rotation and shear \citep{DelSordoBrandenburg2011}, as well as viscosity \citep{MeeBrandenburg2006,FederrathEtAl2011PRL}. Thus, turbulence driven by purely compressive drivers can still contain up to half of its kinetic power in solenoidal modes \citep[][Figure~14]{FederrathDuvalKlessenSchmidtMacLow2010}.

In order to understand the dependence of the dynamo on the compressibility of the plasma, we present in \S\ref{sec:mach} a systematic study in which the turbulent driving and Mach number are varied, covering the subsonic, nearly incompressible regime as well as the high-compressible, supersonic regime of turbulence. We determine the growth rate and saturation level of the magnetic field in both regimes. In \S\ref{sec:pm} we determine the dependence of the compressible turbulent dynamo on the magnetic Prandtl number and the kinematic Reynolds number. 

Many astrophysical systems are characterised by weak or strong magnetic guide fields, i.e., fields that have an ordered mean component along a specified direction. This is in contrast to the turbulent dynamo, where primarily the turbulent un-ordered magnetic field component is amplified on small scales. In order to measure and understand magnetic field amplification in the presence of a strong guide field, we present new simulations in \S\ref{sec:guidefield}, where we systematically measure the strength of the turbulent magnetic field component as a function of increasing magnetic guide field.

Finally, in \S\ref{sec:implications} we discuss the implications of magnetic fields for astrophysical accretion discs, the structure of the interstellar medium, and for star formation.


\section{The Mach number dependence of the turbulent dynamo}  \label{sec:mach}

Here we investigate fundamental properties of the turbulent dynamo---its growth rate and saturation level---in simulations with extremely different driving of the turbulence (solenoidal versus compressive) and with a range of Mach numbers from $\mach=0.02$ to $20$. Some of the results in this section are published in \citet{FederrathEtAl2011PRL}. While these simulations are highly idealised, they serve as a  systematic numerical experiment in which the driving and Mach number are controlled. Turbulent astrophysical systems can often be characterised by a few fundamental numbers that cover the basic physical behaviour. For example, the turbulence in astrophysical accretion discs around young stars is subsonic ($\mach<1$) and is primarily driven by shear or magneto-rotational instability, which drive solenoidal motions (\emph{solenoidal driving}). By contrast, molecular cloud turbulence is highly compressible and supersonic ($\mach>1$), and is driven by a range of physical processes, most of which induce compression (\emph{compressive driving}), such as supernova explosions, expanding radiation fronts from high-mass stellar feedback and/or galactic spiral shocks. For a review of potential drivers of the turbulence in galaxies and the interstellar medium, please see the articles by \citet{Elmegreen2009}, \citet{FederrathDuvalKlessenSchmidtMacLow2010}, \citet{FederrathKlessen2012}, and \citet{PadoanEtAl2014}.

\subsection{Background and open questions}

Most studies of the turbulent dynamo concentrate on incompressible plasmas, with only few studies approaching the effects of compressibility. For example, \citet{HaugenBrandenburgMee2004} obtained critical Reynolds numbers for dynamo action, but did not investigate growth rates or saturation levels, and only studied Mach numbers in the range \mbox{$0.1\leq\mach\leq2.6$}. The energy released by e.g.~supernova explosions, however, drives interstellar and galactic turbulence with Mach numbers of 10--100 \citep{MacLowKlessen2004}. Thus, much higher Mach numbers have to be investigated in order to understand dynamo action in interstellar clouds. It is furthermore tempting to associate such supernova blast waves with compressive driving of turbulence \citep{MeeBrandenburg2006,SchmidtFederrathKlessen2008,FederrathDuvalKlessenSchmidtMacLow2010}. \citet{MeeBrandenburg2006} concluded that it is very hard to excite the turbulent dynamo with this curl-free driving, because it does not directly inject vorticity, $\nabla\times\vect{u}$.

Here we show that the turbulent dynamo can be driven by curl-free driving mechanisms, and we quantify the amplification as a function of compressibility of the plasma. The main questions addressed are: How does the turbulent dynamo depend on the Mach number and driving of the turbulence? What is the field geometry and amplification mechanism? What are the growth rates and saturation levels in the subsonic and supersonic regime?

\subsection{Methods} \label{sec:methods}

\subsubsection{Magnetohydrodynamical equations}

We use a modified version of the FLASH code \citep{FryxellEtAl2000,DubeyEtAl2008} to solve the three-dimensional (3D), compressible, magnetohydrodynamical (MHD) equations on uniform, periodic grids, including viscous and resistive dissipation terms,
\begin{align}
& \ddt\,\rho + \nabla\cdot\left(\rho \bfu\right)=0, \label{eq:mhd1} \\
& \ddt\!\left(\rho \bfu\right) + \nabla\cdot\left(\rho \bfu\!\otimes\!\bfu - \frac{1}{4\pi}\bfB\!\otimes\!\bfB\right) + \nabla p_\mathrm{tot} = \nabla\cdot\left(2\nu\rho\boldsymbol{\mathcal{S}}\right) + \rho{\bf F}, \label{eq:mhd2} \\
& \ddt\,e + \nabla\cdot\left[\left(e+p_\mathrm{tot}\right)\bfu - \frac{1}{4\pi}\left(\bfB\cdot\bfu\right)\bfB\right] = \nabla\cdot\left[2\nu\rho\bfu\cdot\boldsymbol{\mathcal{S}}+\frac{1}{4\pi}\bfB\times\left(\eta\nabla\times\bfB\right)\right], \label{eq:mhd3} \\
& \ddt\,\bfB = \nabla\times\left(\bfu\times\bfB\right) + \eta\nabla^2\bfB, \label{eq:mhd4} \\
& \nabla\cdot\bfB = 0. \label{eq:mhd5}
\end{align}
Here, $\rho$, $\bfu$, $p_\mathrm{tot}=p_\mathrm{th}+ (1/8\pi)\left|\bfB\right|^2$, $\bfB$, and $e=\rho \epsilon_\mathrm{int} + (1/2)\rho\left|\bfu\right|^2 + (1/8\pi)\left|\bfB\right|^2$ denote the gas density, velocity, pressure (thermal plus magnetic), magnetic field, and energy density (internal plus kinetic, plus magnetic), respectively. Physical shear viscosity is included with the traceless rate of strain tensor, $\mathcal{S}_{ij}=(1/2)(\partial_i u_j+\partial_j u_i)-(1/3)\delta_{ij}\nabla\cdot\bfu$ in the momentum equation~(\ref{eq:mhd2}), and controlled by the kinematic viscosity ($\nu$). Physical diffusion of $\bfB$ is controlled by the magnetic diffusivity $\eta=1/(4\pi\sigma)$ (the inverse of the electrical conductivity $\sigma$) in the induction equation~(\ref{eq:mhd4}). The MHD equations are closed with an isothermal equation of state, $p_\mathrm{th}=\cs^2\rho$, where $\cs=\mathrm{const}$ is the sound speed.

We note that equations~(\ref{eq:mhd2}) and (\ref{eq:mhd3}) only contain the shear viscosity $\nu$, while the bulk viscosity is assumed to be zero. For monatomic ideal gases, the bulk viscosity $\xi$ is indeed identically zero, which can be derived from kinetic theory \citep{Mihalas1984}. For polyatomic molecules, this does not need to be the case, but $\xi\neq0$ only, if a relaxation process takes place that is of the same order or slower than a typical fluid timescale \citep{Mihalas1984}. The value of $\xi$ strongly depends on the composition of the polyatomic gas and measurements of $\xi$ from different experiments give different results \citep{Tisza1942}. For simplicity, we set the bulk viscosity $\xi=0$ and only consider the well-understood shear viscosity $\nu$. We further note that using an equation of state for the gas that relates pressure with density and temperature (which is the standard approach in fluid dynamics) implies $\xi=0$ \citep{Truesdell1952}.

\subsubsection{Turbulence driving}

In order to drive turbulence with a target Mach number, we apply the driving field ${\bf F}$ as a source term in the momentum equation~(\ref{eq:mhd2}). The driving field is constructed with a stochastic Ornstein-Uhlenbeck (OU) process \citep{EswaranPope1988,SchmidtEtAl2009,PriceFederrath2010}, implemented by \citet{FederrathDuvalKlessenSchmidtMacLow2010} in the public version of the FLASH code (\url{http://www.flash.uchicago.edu/site/flashcode/}). The OU process yields a driving pattern that varies smoothly in space and time with an auto-correlation time equal to the eddy-turnover time (also called turbulent box-crossing time), $\ted=L/(2\mach\cs)$ on the largest scales ($L/2$) in the periodic simulation domain of side length $L$. The turbulent sonic Mach number is defined as $\mach=u_\mathrm{turb}/\cs$, which is the ratio of the turbulent velocity dispersion $u_\mathrm{turb}$ on scale $L/2$ and the sound speed $\cs$. The driving is constructed in Fourier space such that the kinetic energy is injected at the smallest wave numbers, $1<\left|\mathbf{k}\right|L/2\pi<3$. The peak of energy injection is on scale $L/2$, i.e., $k=2$, and falls off parabolically towards smaller and higher wave numbers such that the driving power reaches zero exactly at $k=1$ and $k=3$. This procedure limits the direct effect of the driving to a narrow wave number band ($1<\left|\mathbf{k}\right|L/2\pi<3$) and allows the turbulence to develop self-consistently on smaller scales ($k\geq3$).

By applying a projection in Fourier space, we can decompose the driving field into its solenoidal and compressive parts. In index notation, the projection operator reads
\begin{equation} \label{eq:projection}
\mathcal{P}_{ij}^\zeta\,(\vect{k}) = \zeta\,\mathcal{P}_{ij}^\perp+(1-\zeta)\,\mathcal{P}_{ij}^\parallel = \zeta\,\delta_{ij}+(1-2\zeta)\,k_i k_j/|\vect{k}|^2,
\end{equation}
where $\mathcal{P}_{ij}^\perp$ and $\mathcal{P}_{ij}^\parallel$ are the solenoidal and compressive projection operators, respectively. This yields a ratio of compressive driving power $F_\mathrm{comp}$ to total driving power $F_\mathrm{tot}$ as a function of the parameter $\zeta$,
\begin{equation} \label{eq:driving_ratio}
\frac{F_\mathrm{comp}}{F_\mathrm{tot}} = \frac{(1-\zeta)^2}{1-2\zeta+3\zeta^2}.
\end{equation}
The projection operation allows us to construct a solenoidal (divergence-free) driving field ($\nabla\cdot\vect{F}=0$) or a compressive (curl-free) driving field ($\nabla\times\vect{F}=0$) by setting $\zeta=1$ (sol) or $\zeta=0$ (comp). Mixed driving ratios are obtained by picking values in the range $1<\zeta<0$. Setting $\zeta=1/2$ yields a ``natural mixture'', i.e., a driving field with $F_\mathrm{comp}/F_\mathrm{tot}=1/3$. The latter is equivalent to the result of randomly-chosen driving modes in 3D, where on average 1 spatial dimension is occupied by longitudinal (compressive) modes and 2 spatial dimensions are occupied by transverse (solenoidal) modes \citep{FederrathKlessenSchmidt2008}.

\subsubsection{Initial conditions and numerical methods}

We start our numerical experiments by setting $L=1.24\times10^{19}\,\cm$, zero initial velocities $\bfu_0=0$, uniform density $\rho_0=1.93\times10^{-21}\,\g\,\cm^{-3}$, $\cs=2\times10^4\,\cm\,\mathrm{s}^{-1}$, and $\bfB=(0,0,B_{0z})$ with $B_{0z}=4.4\times10^{-16}\,\G$ in $z$-direction, corresponding to an extremely high initial plasma $\beta=p_\mathrm{th}/p_\mathrm{m}=10^{20}$. These values are motivated by dynamo studies of primordial clouds \citep{SchleicherEtAl2010,SurEtAl2010,FederrathSurSchleicherBanerjeeKlessen2011,SurEtAl2012}, but in the following, we scale all quantities to dimensionless units to address fundamental questions of magnetic field amplification in compressible plasmas. The physics of these turbulent systems is fully determined by the dimensionless $\mach$ and plasma $\beta$, as well as the driving parameter $\zeta$ (sol vs.~comp). The actual plasma densities, velocities, and magnetic fields can be scaled to astrophysical systems that are described by the same set of basic dimensionless numbers ($\mach$, $\beta$, $\zeta$).

For most of the simulations, we set the kinematic viscosity $\nu$ and the magnetic diffusivity $\eta$ to zero, and thus solve the ideal MHD equations. In this case, the dissipation of kinetic and magnetic energy is caused by the discretisation of the MHD equations. However, we do not add artificial viscosity. We use Riemann solvers, which capture shocks well in the absence of artificial viscosity. In addition to the ideal MHD simulations, we also solve the full, non-ideal MHD equations~(\ref{eq:mhd1})--(\ref{eq:mhd5}) for 4 representative models to show that our results are physical and robust against changes in the numerical scheme. For the ideal MHD simulations, we use the positive-definite, split HLL3R Riemann scheme \citep{WaaganFederrathKlingenberg2011} in FLASH v2.5, while our non-ideal MHD simulations were performed with the un-split staggered-mesh scheme in FLASH v4 \citep{LeeDeane2009}, which uses a third-order reconstruction, constrained transport to maintain $\nabla\cdot\bfB=0$ to machine precision, and the HLLD Riemann solver \citep{MiyoshiKusano2005}. We run simulations with $128^3$, $256^3$, and $512^3$ grid cells, in order to test numerical convergence of our results.

\subsection{Results}

\subsubsection{Time evolution}

\begin{figure*}
\centerline{\includegraphics[width=0.65\linewidth]{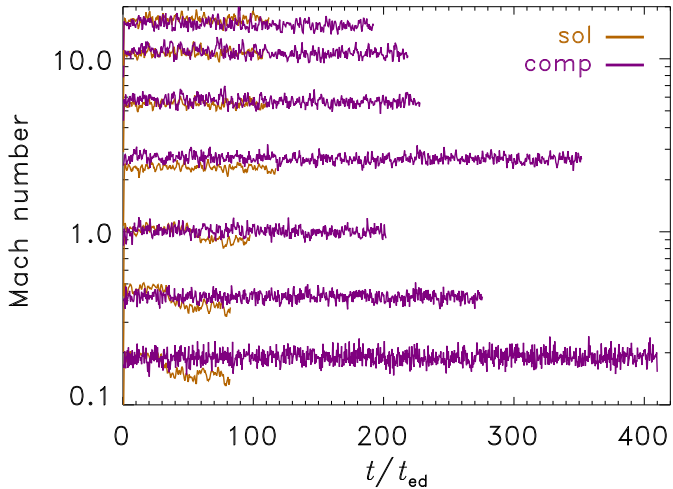}}
\caption{Turbulent sonic Mach number ($\mach$) as a function of the turbulent crossing time ($t/\ted$) for all runs with solenoidal (sol) and compressive (comp) driving. These simulations cover compressible plasmas from subsonic turbulence ($\mach<1$) up into the highly compressible, supersonic regime ($\mach>1$).}
\label{fig:prl_mach}
\end{figure*}

The turbulence is fully developed after an initial transient phase that lasts for $2\,\ted$ \citep[e.g.,][]{KitsionasEtAl2009,FederrathKlessenSchmidt2009,PriceFederrath2010}, and the Mach number reaches its target value, fluctuating on a 10\% level. Figure~\ref{fig:prl_mach} shows the time evolution of $\mach$ in all runs with solenoidal and compressive driving of the turbulence. Some simulations had to be run for a few hundred crossing times in order to reach saturation of the magnetic field. This is the case in all the compressively driven simulations, which---as we will see below---have significantly lower dynamo growth rates than in the case of solenoidal driving. Interesting to note is the drop in $\mach$ for the solenoidally driven runs with $\mach\lesssim1$ towards late times, $t\gtrsim50\,\ted$. This slight drop in Mach number indicates that the magnetic field has reached very high saturation levels, such that the back reaction of the magnetic field via the Lorentz force has become significant in these models. However, the magnetic field has little dynamical impact on the turbulent flow in all supersonic runs and in all runs with compressive driving. Although the kinematics of the turbulent flow is not strongly affected in those cases, the structure and fragmentation of the gas is still influenced by the presence of a turbulent magnetic field \citep{HennebelleTeyssier2008,FederrathKlessen2012,PadoanEtAl2014}.

\begin{figure}
\centerline{\includegraphics[width=0.70\linewidth]{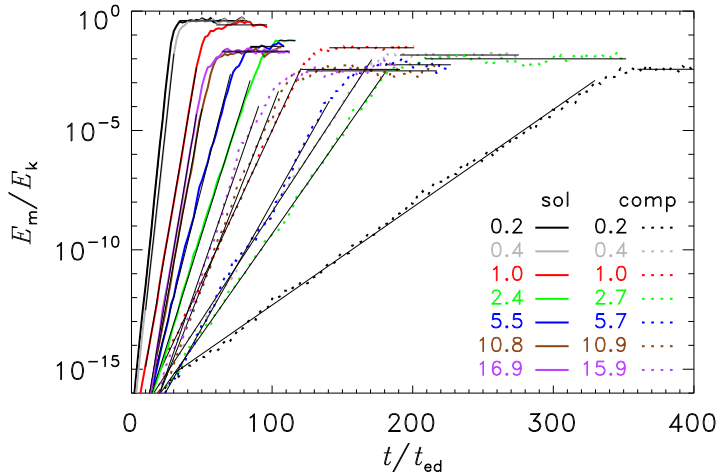}}
\caption{Magnetic-to-kinetic energy ratio ($E_\mathrm{m}/E_\mathrm{k}$) as a function of the turbulent crossing time ($t/\ted$) for all runs with solenoidal (sol) and compressive (comp) driving. The time-averaged sonic Mach number ($\mach$) of each model is indicated in the legend (see figure~\ref{fig:prl_mach} for the time evolution of $\mach$). The thin solid lines show exponential fits in the regime of turbulent dynamo amplification, followed by constant fits in the saturation phase. The evolution of $E_\mathrm{m}/E_\mathrm{k}$ reveals extremely different efficiencies of the dynamo, depending on the Mach number and driving of the turbulence.}
\label{fig:prl_evol}
\end{figure}

Figure~\ref{fig:prl_evol} shows the ratio of magnetic-to-kinetic energy ($E_\mathrm{m}/E_\mathrm{k}$) as a function of time. We immediately see that the magnetic energy grows exponentially over several orders of magnitude in any of the simulation models and finally reaches saturation at different levels (indicated by the fitted constant line towards late times when the curves saturate).

\subsubsection{Density and magnetic field structure}

\begin{figure*}
\centerline{\includegraphics[width=0.95\linewidth]{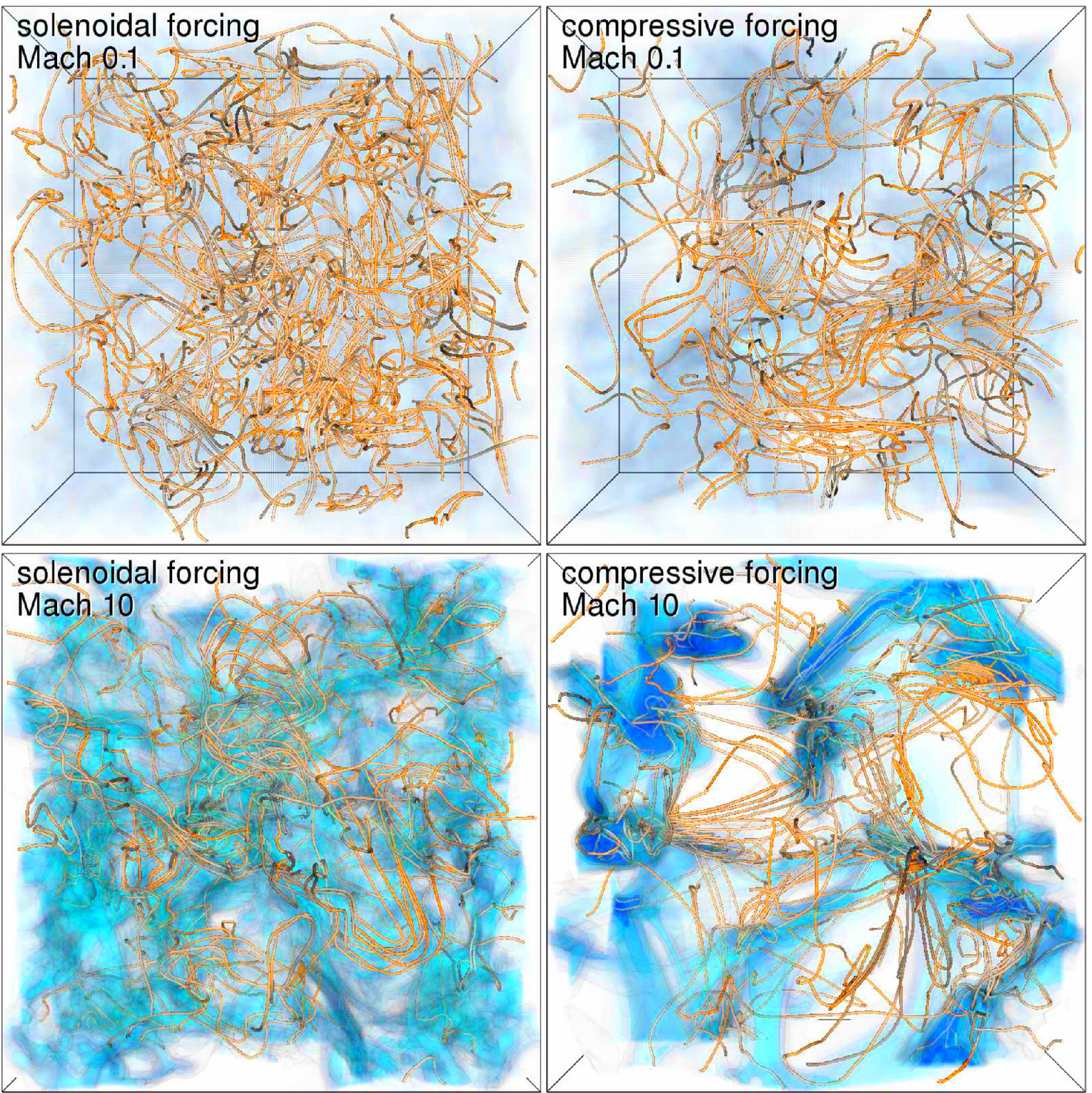}}
\caption{3D renderings of the gas density contrast on a logarithmic scale \mbox{($0.5\leq\rho/\rho_0\leq50$)} (from white to blue), and magnetic field lines (orange) for solenoidal driving at \mbox{$\mach=0.1$} (top left) and \mbox{$\mach=10$} (bottom left), and compressive driving at \mbox{$\mach=0.1$} (top right) and \mbox{$\mach=10$} (bottom right). The stretch-twist-fold mechanism of the dynamo \citep{BrandenburgSubramanian2005} is evident in all models, but operates with different efficiency due to the varying compressibility, flow structure, and formation of shocks in the supersonic plasmas. From \citet{FederrathEtAl2011PRL}. \emph{An animation is available at \url{http://www.mso.anu.edu.au/~chfeder/pubs/dynamo_prl/dynamo_prl.html}.}}
\label{fig:prl_snapshots}
\end{figure*}

Figure~\ref{fig:prl_snapshots} shows 3D volume renderings of some the extreme cases (solenoidal driving on the left and compressive driving on the right, each for $\mach=0.1$ in the top panels and $\mach=10$ in the bottom panels). We see that the supersonic simulation runs (bottom panels) are dominated by shocks. Compressive driving yields stronger density enhancements for similar Mach numbers compared to solenoidal driving \citep{NordlundPadoan1999,FederrathKlessenSchmidt2008,FederrathDuvalKlessenSchmidtMacLow2010,PriceFederrathBrunt2011,KonstandinEtAl2012ApJ,FederrathKlessen2013,GinsburgFederrathDarling2013,KainulainenFederrathHenning2013,Federrath2013}. The magnetic field occupies large volume fractions with relatively straight field lines in the compressively driven cases, while solenoidal driving produces more space-filling, tangled field configurations. This suggests that the turbulent dynamo operates more efficiently in the case of solenoidal driving, which is quantified and explained in the following.

\subsubsection{Dynamo growth rate and saturation level}

\begin{figure}
\centerline{\includegraphics[width=0.65\linewidth]{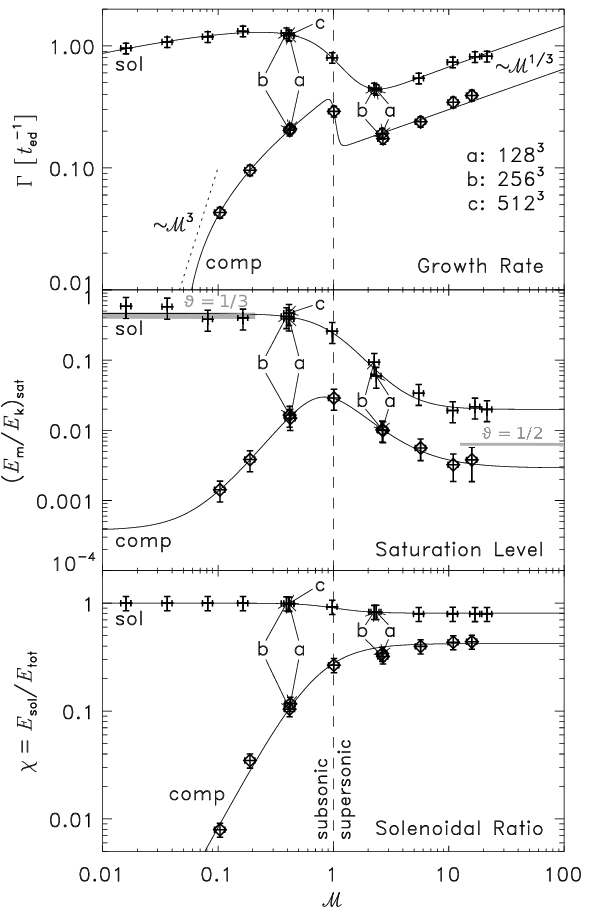}}
\caption{Growth rate (top), saturation level (middle), and solenoidal ratio (bottom) as a function of Mach number, for all simulations with solenoidal (crosses) and compressive driving (diamonds). The thin solid lines show empirical fits with equation~(\ref{eq:fit_prl}), the parameters of which are listed in table~\ref{tab:fit}. The arrows point to 4 simulations ($\mach\approx0.4$, $2.5$ for sol.~and comp.~driving), which used ideal MHD on $128^3$ grid cells (a), non-ideal MHD on $256^3$ (b), and $512^3$ grid cells (c), demonstrating convergence for the given magnetic Prandtl number, $\Pm\approx2$, and kinematic Reynolds number, $\mathrm{Re}\approx1500$. The theoretical predictions for the saturation level from \citet{SchoberEtAl2015} are added as grey lines (middle panel) in the limit $\mach\ll1$ (Kolmogorov scaling exponent: $\vartheta=1/3$) and $\mach\gg1$ (Burgers scaling exponent: $\vartheta=1/2$).}
\label{fig:prl_gratesat}
\end{figure}

\begin{table}
\begin{center}
\begin{tabular}{l@{\hskip 0.8cm}cc@{\hskip 0.5cm}cc@{\hskip 0.5cm}cc}
  & \multicolumn{2}{c@{\hskip 0.5cm}}{Growth rate $\Gamma\,\left[\ted^{-1}\right]$} & \multicolumn{2}{c@{\hskip 0.5cm}}{Saturation level $\satlev$} & \multicolumn{2}{c}{Solenoidal ratio $\solratio$} \\[2pt]
  & (sol) & (comp) & (sol) & (comp) & (sol) & (comp) \\[3pt]
$p_0$ & -18.71                     & \phantom{-}2.251 & \phantom{-}0.020 & \phantom{-}0.037 & \phantom{-}0.808 & \phantom{-}0.423 \\
$p_1$ & \phantom{-}0.051 & \phantom{-}0.119 & \phantom{-}2.340 & \phantom{-}1.982 & \phantom{-}2.850 & \phantom{-}1.970 \\
$p_2$ & -1.059                     & -0.802                     & \phantom{-}23.33 & -0.027                    & \phantom{-}1.238 & 0 \\
$p_3$ & \phantom{-}2.921 & \phantom{-}25.53 & \phantom{-}2.340 & \phantom{-}3.601 & \phantom{-}2.850 & \phantom{-}1.970 \\
$p_4$ & \phantom{-}1.350 & \phantom{-}1.686 & 1                              & \phantom{-}0.395 & 1                              & \phantom{-}0.535 \\
$p_5$ & \phantom{-}0.313 & \phantom{-}0.139 & 0                              & \phantom{-}0.003 & 0                              & 0 \\
$p_6$ & 1/3                           & 1/3                          & 0                              & 0                              & 0                              & 0 
\end{tabular}
\caption{Parameters in equation~(\ref{eq:fit_prl}) for the fits in figure~\ref{fig:prl_gratesat}.}
\label{tab:fit}
\end{center}
\end{table}

Figure~\ref{fig:prl_gratesat} (top and middle panels) show the dynamo growth rate ($\Gamma$) in the relation $\Em=\Emzero\exp(\Gamma t)$ and the saturation level $\satlev$ as a function of Mach number for all simulation models. Both $\Gamma$ and $\satlev$ depend on $\mach$ and on the driving of the turbulence. Solenoidal driving yields growth rates and saturation levels higher than for compressive driving. Both $\Gamma$ and $\satlev$ change significantly at the transition from subsonic to supersonic turbulence. Within a turbulent medium, the transition from supersonic to subsonic turbulence occurs on the sonic scale \citep{VazquezBallesterosKlessen2003,FederrathKlessen2012,Hopkins2013IMF,Federrath2016}. We conclude that the formation of shocks at $\mach\approx1$ is responsible for destroying some of the coherent vortex motions necessary to drive the dynamo \citep{HaugenBrandenburgMee2004}. However, as $\mach$ is increased further, vorticity generation in oblique, colliding shocks \citep{SunTakayama2003,KritsukEtAl2007} starts to dominate over the destruction of vorticity.

The data points labelled with `a', `b', and `c' in figure~\ref{fig:prl_gratesat} show simulations with different numerical resolution ($128^3$, $256^3$ and $512^3$ cells, respectively) and with different numerical solvers/schemes (`a': HLL3R with $\nu$ and $\eta$ set to zero, i.e., ideal MHD; `b' and `c': HLLD with non-ideal MHD; cf.~\S\ref{sec:methods}). These convergence studies demonstrate that our results do not depend on the choice of numerical solver and, in particular, do not depend on the choice of dissipation mechanism \citep[see also][]{BenziEtAl1993}. The reason for this is that the physical dissipation was chosen to be similar to the numerical one in the low-resolution simulations, while in the high-resolution simulations, the physical dissipation dominates the numerical dissipation. This is why our simulations are converged and the results do not depend on the choice of numerical solver.

The very small growth rates in the subsonic, compressively driven models are due to the fact that hardly any vorticity is generated. In order to quantify this, we show the solenoidal ratio, i.e., the specific kinetic energy in solenoidal modes of the turbulent velocity field divided by the total specific kinetic energy, $\chi=\solratio$ in the bottom panel of figure~\ref{fig:prl_gratesat}. We see that the solenoidal ratio drops sharply towards lower Mach numbers in the case of compressive driving. In the absence of the baroclinic term, $(1/\rho^2)\nabla\rho\times\nabla p_\mathrm{th}$, the only way to generate vorticity, $\bfomega=\nabla\times\bfu$, with compressive (curl-free) driving is via viscous interactions in the vorticity equation \citep{MeeBrandenburg2006},
\begin{equation}
\partial_t \bfomega = \nabla\times\left(\bfu\times\bfomega\right) + \nu\nabla^2\bfomega + 2\nu\nabla\times\left(\boldsymbol{\mathcal{S}}\nabla\ln\rho\right).
\label{eq:vorticity}
\end{equation}
While the second term on the right hand side is diffusive and dampens $\bfomega$, the last term can actually generate vorticity through viscous interactions in the presence of logarithmic density gradients. Thus, even in the absence of initial vorticity, small seeds of $\bfomega$ can be generated. Those are then exponentially amplified by the non-linear term, $\nabla\times\left(\bfu\times\bfomega\right)$, if the Reynolds numbers are sufficiently large \citep{Frisch1995}\footnote{The general mechanism for vorticity amplification is thus analogous to the dynamo amplification of small seed fields, because of the similar form of equations~(\ref{eq:vorticity}) and~(\ref{eq:mhd4}).}. For very low Mach numbers, however, density gradients start to vanish, thus explaining the steep drop of the dynamo growth rate for compressively driven turbulence with low Mach numbers.

Analytic estimates by \citet{MossShukurov1996} suggest that the dynamo growth rate $\Gamma\propto\mach^3$ in compressively driven, acoustic turbulence, shown as the dotted line in the top panel of figure~\ref{fig:prl_gratesat} (we note that we defined $\ted=L/(2\mach\cs)$, while \citet{MossShukurov1996} defined $\ted=L/(2\cs)$, which differs by a factor of $\mach$ from our definition). Our simulations are consistent with this, though simulations at even lower Mach number for compressive driving are required to confirm the $\Gamma\propto\mach^3$ scaling.

The solid lines in figure~\ref{fig:prl_gratesat} show fits with the empirical function,
\begin{equation}
\label{eq:fit_prl}
f(x)=\left(p_0\,\frac{x^{p_1}+p_2}{x^{p_3}+p_4}+p_5\right)x^{p_6}.
\end{equation}
The fit parameters are given in table~\ref{tab:fit}. We caution that these fits do not necessarily reflect the true asymptotic behaviour of $\Gamma$ and $\satlev$, but they provide a reasonably good fit for the range of Mach numbers tested in our simulations.

The subsonic, solenoidally driven models show very high saturation levels, $\satlev\approx40$--60\%, in excellent agreement with the theoretical predictions from \citet{SchoberEtAl2015} for \citet{Kolmogorov1941c} turbulence (velocity scaling exponent $\vartheta=1/3$). On the other hand, for the highly compressible, supersonic limit, \citet{SchoberEtAl2015} predict $\sim\!100$ times lower saturation levels, of the order of $0.6\%$, which is qualitatively consistent with our MHD simulations. The prediction by \citet{SchoberEtAl2015} for $\mach\gg1$ is based on the scaling exponent of supersonic \citet{Burgers1948} turbulence, $\vartheta=1/2$, which is indeed measured in numerical simulations \citep{KritsukEtAl2007,FederrathDuvalKlessenSchmidtMacLow2010,Federrath2013}.

The high saturation levels in the case of subsonic, solenoidally driven turbulence cause a strong back reaction of the magnetic field, explaining the Mach number drop in the saturation regime that we saw in figure~\ref{fig:prl_mach} for these models. This is consistent with the simulations in \citet{HaugenBrandenburgDobler2003,HaugenBrandenburgDobler2004}. For the growth rate, we fix $p_6$ such that $\Gamma\propto\mach^{1/3}$ for $\mach\gg1$, in good agreement with our models up to $\mach\approx20$. However, simulations with even higher $\mach$ are required to see if $\Gamma\propto\mach^{1/3}$ holds in this limit. We find that $\Gamma$ depends much less on $\mach$ in the case of solenoidal driving. Nevertheless, a drop of the growth rate at $\mach\approx1$ is present in both driving cases.

Dynamo theories based on the phenomenological model of incompressible, purely solenoidal turbulence by \citet{Kolmogorov1941c} predict no dependence of $\Gamma$ on $\mach$. For instance, \citet{Subramanian1997} derived $\Gamma=(15/24)\mathrm{Re}^{1/2}\ted^{-1}$ based on the Kolmogorov-Fokker-Planck equation, in the limit of large magnetic Prandtl number, $\Pm=\nu/\eta=\Rm/\mathrm{Re}\gg1$ with the kinetic and magnetic Reynolds numbers $\mathrm{Re}=L u_\mathrm{turb}/(2\nu)$ and $\Rm=L u_\mathrm{turb}/(2\eta)$. For $\Pm\approx2$ \citep[which is typically the result of numerical dissipation in the ideal MHD approximation; see][]{LesaffreBalbus2007}, and $\mathrm{Re}\approx1500$, corresponding to our simulations, we find slightly smaller growth rates than predicted in \citet{Subramanian1997}, but in agreement with analytic considerations \citep{BoldyrevCattaneo2004}, and with numerical simulations of incompressible turbulence for $\Pm\approx1$ \citep{SchekochihinEtAl2007,ChoEtAl2009}. Thus, an extension of the dynamo theory to small $\Pm$ is needed and will be presented in \S\ref{sec:pm} below. Moreover, extending the theory from Kolmogorov to Burgers, shock-dominated turbulence is a necessary next step in order to develop a more generalised theory of turbulent dynamos, potentially with predictive power for the supersonic, highly compressible regime.

In summary, we find that the growth rate and saturation level of the turbulent dynamo strongly depend on the Mach number and the driving of the turbulence. We conclude that the compressibility of the plasma plays a crucial role for the amplification of turbulent magnetic fields.


\section{Dependence of the dynamo on the magnetic Prandtl number}  \label{sec:pm}

Previous studies of incompressible turbulence have demonstrated that the turbulent dynamo depends on the magnetic Prandtl number, $\pmag=\nu/\eta=\rmag/\mathrm{Re}$, defined as the ratio of kinematic viscosity $\nu$ to magnetic diffusivity $\eta$ \citep{SchekochihinEtAl2004}. The magnetic Prandtl number also plays a role for the magneto-rotational instability and the transport of angular momentum in accretion disks \citep{Fromang2010,FromangEtAl2010}.

Here we extend this study of $\pmag$ into the compressible, supersonic regime of turbulence. Some of the results discussed in this section are published in \citet{FederrathSchoberBovinoSchleicher2014}.

On large cosmological scales and in the interstellar medium, we typically have $\pmag\gg1$ \citep[for details on how to estimate the viscosity and magnetic diffusivity, see][for example, in molecular clouds, we have $\mathrm{Re}\approx10^7$ and $\Rm\approx10^{17}$, hence $\pmag\approx10^{10}$]{WardleNg1999,PintoGalli2008,SchoberEtAl2012,Krumholz2014}, while for the interior of stars and planets, the case with $\pmag\ll1$ is more relevant \citep{SchekochihinEtAl2007}. However, numerical simulations are typically restricted to $\pmag\approx1$, because of limited scale separation that can be achieved with currently accessible numerical resolutions. Simulations by \citet{IskakovEtAl2007} have demonstrated that the turbulent dynamo operates for $\pmag\lesssim1$ in incompressible gases, even though an asymptotic scaling relation could not be confirmed. While the bulk of previous work was dedicated to exploring the turbulent dynamo for incompressible plasmas \citep{BrandenburgSokoloffSubramanian2012}, most astrophysical systems show signs of high compressibility. This is especially true for the early Universe when the first stars and galaxies formed \citep{LatifSchleicherSchmidt2014}, in the interstellar medium of present-day galaxies \citep{Larson1981}, as well as in the intergalactic medium \citep{VazzaEtAl2009,IapichinoVielBorgani2013,MiniatiBeresnyak2015}. As in \S\ref{sec:mach}, the compressibility of the plasma is characterised in terms of the sonic Mach number $\mach$.

Based on the Kazantsev model \citep{Kazantsev1968,KazantsevEtAl1985} (see also recent work by \citet{XuLazarian2016}), \citet{SchoberEtAl2012PRE2} derived an analytical dynamo solution for the limiting cases $\pmag\to \infty$ and $\pmag\to0$, considering different scaling relations of the turbulence. \citet{BovinoEtAl2013} later computed numerical solutions of the Kazantsev equation for intermediate values of $\pmag$. These theoretical studies suggested that the turbulent dynamo operates for a range of $\pmag$, as long as the magnetic Reynolds number is sufficiently high. A central restriction of the Kazantsev model is the assumption of an incompressible velocity field, for which a separation into solenoidal and compressible parts is not necessary. But the distinction between solenoidal and compressible modes is essential for highly compressible, supersonic turbulence (see~\S\ref{sec:mach}). Moreover, the Kazantsev framework assumes that the turbulence is $\delta$-correlated in time, but the resulting uncertainties introduced by this assumption are only a few per cent \citep{SchekochihinKulsrud2001,KleeorinRogachevskiiSokoloff2002,BhatSubramanian2014}. However, the assumption of incompressibility may introduce significantly higher uncertainties in the analytic and semi-analytic estimates based on the Kazantsev model. Ultimately one needs full 3D simulations to determine the dependence of the dynamo growth rates and saturation levels on $\pmag$.

Here we present a systematic study of the turbulent dynamo and its dependence on the magnetic Prandtl number in the highly compressible regime. For this purpose, we consider supersonic turbulence with $\mach=4$--$11$, and magnetic Prandtl numbers between $\pmag=0.1$ and $10$. The simulation results are compared with analytic and semi-analytic predictions based on the Kazantsev model.

\subsection{Numerical simulations} \label{sec:sims}

The simulations follow the same basic approach as explained in detail in \S\ref{sec:methods}. We run most of the simulations until saturation of the magnetic field is reached. We determine the growth rate by fitting an exponential function to the time evolution of the magnetic energy. The saturation level is determined by measuring the magnetic-to-kinetic energy ratio $\esat$ in the saturated phase. We note that the turbulent dynamo has also been studied with ``shell models'' \citep[][and references therein]{FrickEtAl2006}. Shell models can also provide theoretical predictions for the magnetic energy growth in the exponential and saturated regimes of the dynamo, and they are complementary to the 3D numerical simulations presented here.

\begin{table}
\begin{center}
\begin{tabular*}{\linewidth}{@{\extracolsep{\fill} }lccccccc}
Simulation Model & $N_\mathrm{res}^3$ & $\mach$ & $\pmag$ & $\re$ & $\rmag$ & $\Gamma\;(\ted^{-1})$ & $\esat$ \\[3pt]
\texttt{Dyn\_512\_Pm0.1\_Re1600} & $ 512^3$ & $ 11$ & $0.1$ & $ 1600$ & $  160$ & $(  2.7\pm3.0)\!\times\!10^{-3 }$ & n/a \\
\texttt{Dyn\_1024\_Pm0.1\_Re1600} & $1024^3$ & $ 11$ & $0.1$ & $ 1600$ & $  160$ & $(  1.9\pm 50)\!\times\!10^{-3}\;$ & n/a \\
\texttt{Dyn\_512\_Pm0.2\_Re1600} & $ 512^3$ & $ 11$ & $0.2$ & $ 1600$ & $  320$ & $(  3.5\pm0.4)\!\times\!10^{-2 }$ & $(  6.0\pm2.0)\!\times\!10^{-4 }$ \\
\texttt{Dyn\_512\_Pm0.5\_Re1600} & $ 512^3$ & $ 11$ & $0.5$ & $ 1600$ & $  810$ & $(  2.0\pm0.2)\!\times\!10^{-1 }$ & $(  1.0\pm0.3)\!\times\!10^{-2 }$ \\
\texttt{Dyn\_512\_Pm2\_Re1600} & $ 512^3$ & $ 11$ & $  2$ & $ 1600$ & $ 3200$ & $(  4.5\pm0.4)\!\times\!10^{-1 }$ & $(  3.0\pm1.0)\!\times\!10^{-2 }$ \\
\texttt{Dyn\_256\_Pm5\_Re1600} & $ 256^3$ & $ 11$ & $  5$ & $ 1600$ & $ 8000$ & $(  6.4\pm0.6)\!\times\!10^{-1 }$ & $(  3.9\pm1.3)\!\times\!10^{-2 }$ \\
\texttt{Dyn\_512\_Pm5\_Re1600} & $ 512^3$ & $ 11$ & $  5$ & $ 1600$ & $ 8000$ & $(  5.8\pm0.6)\!\times\!10^{-1 }$ & $(  5.2\pm1.7)\!\times\!10^{-2 }$ \\
\texttt{Dyn\_1024\_Pm5\_Re1600} & $1024^3$ & $ 11$ & $  5$ & $ 1600$ & $ 8000$ & $(  6.2\pm0.6)\!\times\!10^{-1 }$ & $(  4.6\pm1.5)\!\times\!10^{-2 }$ \\
\texttt{Dyn\_256\_Pm10\_Re1600} & $ 256^3$ & $ 11$ & $ 10$ & $ 1600$ & $16000$ & $(  6.9\pm0.7)\!\times\!10^{-1 }$ & $(  4.0\pm1.3)\!\times\!10^{-2 }$ \\
\texttt{Dyn\_512\_Pm10\_Re1600} & $ 512^3$ & $ 11$ & $ 10$ & $ 1600$ & $16000$ & $(  6.4\pm0.6)\!\times\!10^{-1 }$ & $(  5.7\pm1.9)\!\times\!10^{-2 }$ \\
\texttt{Dyn\_1024\_Pm10\_Re1600} & $1024^3$ & $ 11$ & $ 10$ & $ 1600$ & $16000$ & $(  6.5\pm0.6)\!\times\!10^{-1 }$ & $(  4.8\pm1.6)\!\times\!10^{-2 }$ \\
\texttt{Dyn\_128\_Pm10\_Re4.7} & $ 128^3$ & $4.0$ & $ 10$ & $  4.7$ & $   47$ & $\;(  6.0\pm160)\!\times\!10^{-3 }$ & n/a \\
\texttt{Dyn\_256\_Pm10\_Re4.6} & $ 256^3$ & $3.9$ & $ 10$ & $  4.6$ & $   46$ & $\;(  5.8\pm160)\!\times\!10^{-3 }$ & n/a \\
\texttt{Dyn\_128\_Pm10\_Re15} & $ 128^3$ & $6.4$ & $ 10$ & $   15$ & $  150$ & $(  3.4\pm0.6)\!\times\!10^{-2 }$ & n/a \\
\texttt{Dyn\_256\_Pm10\_Re15} & $ 256^3$ & $6.4$ & $ 10$ & $   15$ & $  150$ & $(  4.3\pm0.7)\!\times\!10^{-2 }$ & n/a \\
\texttt{Dyn\_128\_Pm10\_Re26} & $ 128^3$ & $7.5$ & $ 10$ & $   26$ & $  260$ & $(  2.9\pm0.3)\!\times\!10^{-1 }$ & $(  4.1\pm1.6)\!\times\!10^{-2 }$ \\
\texttt{Dyn\_256\_Pm10\_Re26} & $ 256^3$ & $7.6$ & $ 10$ & $   26$ & $  260$ & $(  2.6\pm0.3)\!\times\!10^{-1 }$ & $(  4.3\pm1.4)\!\times\!10^{-2 }$ \\
\texttt{Dyn\_128\_Pm10\_Re39} & $ 128^3$ & $8.2$ & $ 10$ & $   39$ & $  390$ & $(  3.4\pm0.3)\!\times\!10^{-1 }$ & $(  4.3\pm1.4)\!\times\!10^{-2 }$ \\
\texttt{Dyn\_256\_Pm10\_Re38} & $ 256^3$ & $8.2$ & $ 10$ & $   38$ & $  380$ & $(  3.2\pm0.3)\!\times\!10^{-1 }$ & $(  5.0\pm1.9)\!\times\!10^{-2 }$ \\
\texttt{Dyn\_512\_Pm10\_Re38} & $ 512^3$ & $8.2$ & $ 10$ & $   38$ & $  380$ & $(  3.1\pm0.6)\!\times\!10^{-1 }$ & n/a \\
\texttt{Dyn\_512\_Pm10\_Re88} & $ 512^3$ & $9.4$ & $ 10$ & $   88$ & $  880$ & $(  4.5\pm0.5)\!\times\!10^{-1 }$ & $(  5.2\pm1.7)\!\times\!10^{-2 }$ \\
\texttt{Dyn\_512\_Pm10\_Re190} & $ 512^3$ & $ 10$ & $ 10$ & $  190$ & $ 1900$ & $(  5.4\pm0.5)\!\times\!10^{-1 }$ & $(  6.0\pm2.0)\!\times\!10^{-2 }$ \\
\texttt{Dyn\_512\_Pm10\_Re390} & $ 512^3$ & $ 10$ & $ 10$ & $  390$ & $ 3900$ & $(  5.9\pm0.6)\!\times\!10^{-1 }$ & $(  6.3\pm2.1)\!\times\!10^{-2 }$ \\
\texttt{Dyn\_256\_Pm10\_Re790} & $ 256^3$ & $ 10$ & $ 10$ & $  790$ & $ 7900$ & $(  6.5\pm0.6)\!\times\!10^{-1 }$ & $(  5.3\pm1.8)\!\times\!10^{-2 }$ \\
\texttt{Dyn\_512\_Pm10\_Re790} & $ 512^3$ & $ 11$ & $ 10$ & $  790$ & $ 7900$ & $(  6.6\pm0.7)\!\times\!10^{-1 }$ & $(  6.4\pm2.1)\!\times\!10^{-2 }$ \\
\end{tabular*}
\caption{Turbulent dynamo simulations with different magnetic Prandtl number ($\pmag$).}
\label{tab:sims}
\end{center}
\end{table}

Here we study the dependence of the turbulent dynamo on $\pmag$, which is accomplished by varying the physical viscosity and magnetic diffusivity. Table~\ref{tab:sims} provides a complete list of all simulations and key parameters. The MHD equations~(\ref{eq:mhd1})--(\ref{eq:mhd5}) are solved with the positive-definite second-order accurate HLL3R Riemann scheme \citep{WaaganFederrathKlingenberg2011}. As a numerical convergence test, we run simulations with \mbox{$N_\mathrm{res}^3=128^3$--$1024^3$} grid points.

\subsection{Dynamo theory} \label{sec:kaztheory}

Most theoretical models of the turbulent dynamo are based on the \emph{Kazantsev} framework \citep{Kazantsev1968,Subramanian1997,RogachevskiiKleeorin1997,BrandenburgSubramanian2005},
The Kazantsev equation assumes zero helicity, $\delta$-correlation in time, and does not take into account the mixture of solenoidal-to-compressible modes in the turbulent velocity field. These limitations are related to the fact that the Kazantsev equation was historically only applied to incompressible turbulence. Here we present an extension of the Kazantsev model to compressible, supersonic regime of turbulence \citep[the details of which are published in][]{SchoberEtAl2012PRE,SchoberEtAl2012PRE2,BovinoEtAl2013,SchleicherEtAl2013,SchoberEtAl2015}.

The form of the Kazantsev equation is similar to the Schr\"odinger equation. This allows us to solve the equation both numerically and analytically with standard methods from quantum mechanics such as the Wentzel-Kramers-Brillouin (WKB) approximation. In order find a solution, we have to assume a scaling of the turbulent velocity fluctuations $u_\mathrm{turb}(\ell)$ in the inertial range ($\ell_\nu < \ell < L$), 
\begin{equation}
u_\mathrm{turb}(\ell) \propto \ell^{\vartheta},
\label{eq:v_inertial}
\end{equation}
where $\ell_\nu$ and $L$ are the viscous and integral scales, respectively. The exponent $\vartheta$ varies from $1/3$ for incompressible, non-intermittent \citet{Kolmogorov1941c} turbulence up to $1/2$ for supersonic, shock-dominated turbulence \citep{Burgers1948}. Based on numerical simulations of mildly supersonic turbulence with Mach numbers \mbox{$\mach\approx2$--$7$}, scaling exponents \mbox{$\vartheta\approx0.37$--$0.47$} were found in numerical simulations \citep{BoldyrevNordlundPadoan2002b,KowalLazarian2010,FederrathDuvalKlessenSchmidtMacLow2010}. Highly supersonic turbulence with $\mach>15$ asymptotically approaches the Burgers limit, $\vartheta=0.5$ \citep{Federrath2013}. Observations of interstellar clouds suggest similar velocity scaling exponents with \mbox{$\vartheta\approx0.38$--$0.5$} \citep{Larson1981,SolomonEtAl1987,OssenkopfMacLow2002,HeyerBrunt2004,RomanDuvalEtAl2011}. Given this range of turbulence scaling exponents from numerical simulations and observations, we here compute theoretical estimates of the dynamo growth rate for $\vartheta=0.35$, $0.40$, and $0.45$.

\subsection{Results and discussion} \label{sec:results}

\subsubsection{Magnetic field structure in low-$\pmag$ and high-$\pmag$ supersonic plasmas}

\begin{figure*}
\centerline{\includegraphics[width=1.0\linewidth]{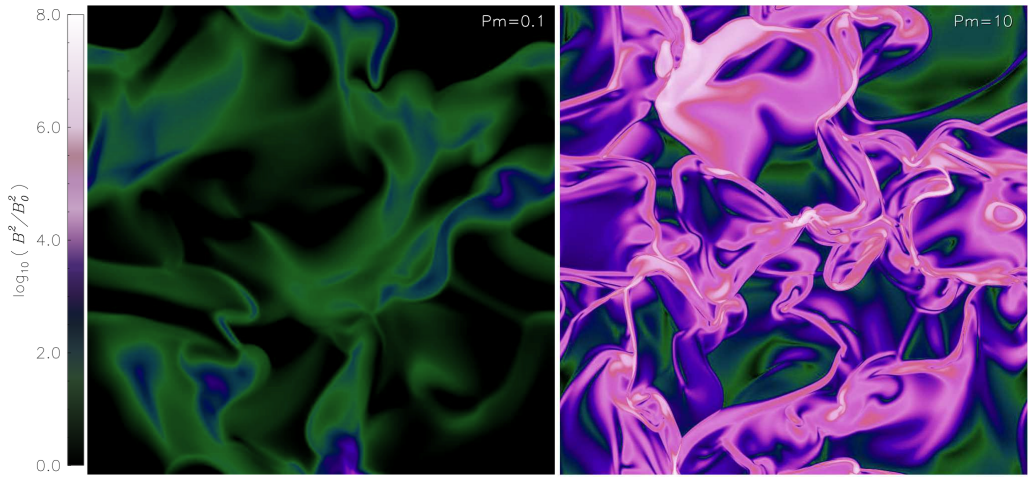}}
\caption{Magnetic energy slices through the mid plane of our dynamo simulations with grid resolutions of $1024^3$ points. The magnetic field grows more slowly for low magnetic Prandtl number $\pmag=0.1$ (\emph{left-hand panel}) compared to $\pmag=10$ (\emph{right-hand panel}). However, dynamo action occurs in both cases, and for the first time shown in highly compressible, supersonic plasmas \citep{FederrathSchoberBovinoSchleicher2014}. \emph{An animation is available at \url{http://www.mso.anu.edu.au/~chfeder/pubs/dynamo_pm/dynamo_pm.html}.}}
\label{fig:images}
\end{figure*}

Figure~\ref{fig:images} provides a visual comparison of the differences in magnetic energy between low-$\pmag$ and high-$\pmag$ supersonic plasmas. Magnetic dissipation is stronger in low-$\pmag$ compared to high-$\pmag$ plasmas (for $\re=\mathrm{const}$). Here we find that the dynamo operates in both cases, but with extremely different efficiency. This is the first time that dynamo action was confirmed in $\pmag<1$, highly compressible, supersonic plasmas \citep{FederrathSchoberBovinoSchleicher2014}.

\subsubsection{Growth rate and saturation level as a function of $\pmag$ and $\re$}

\begin{figure*}
\centerline{\includegraphics[width=1.0\linewidth]{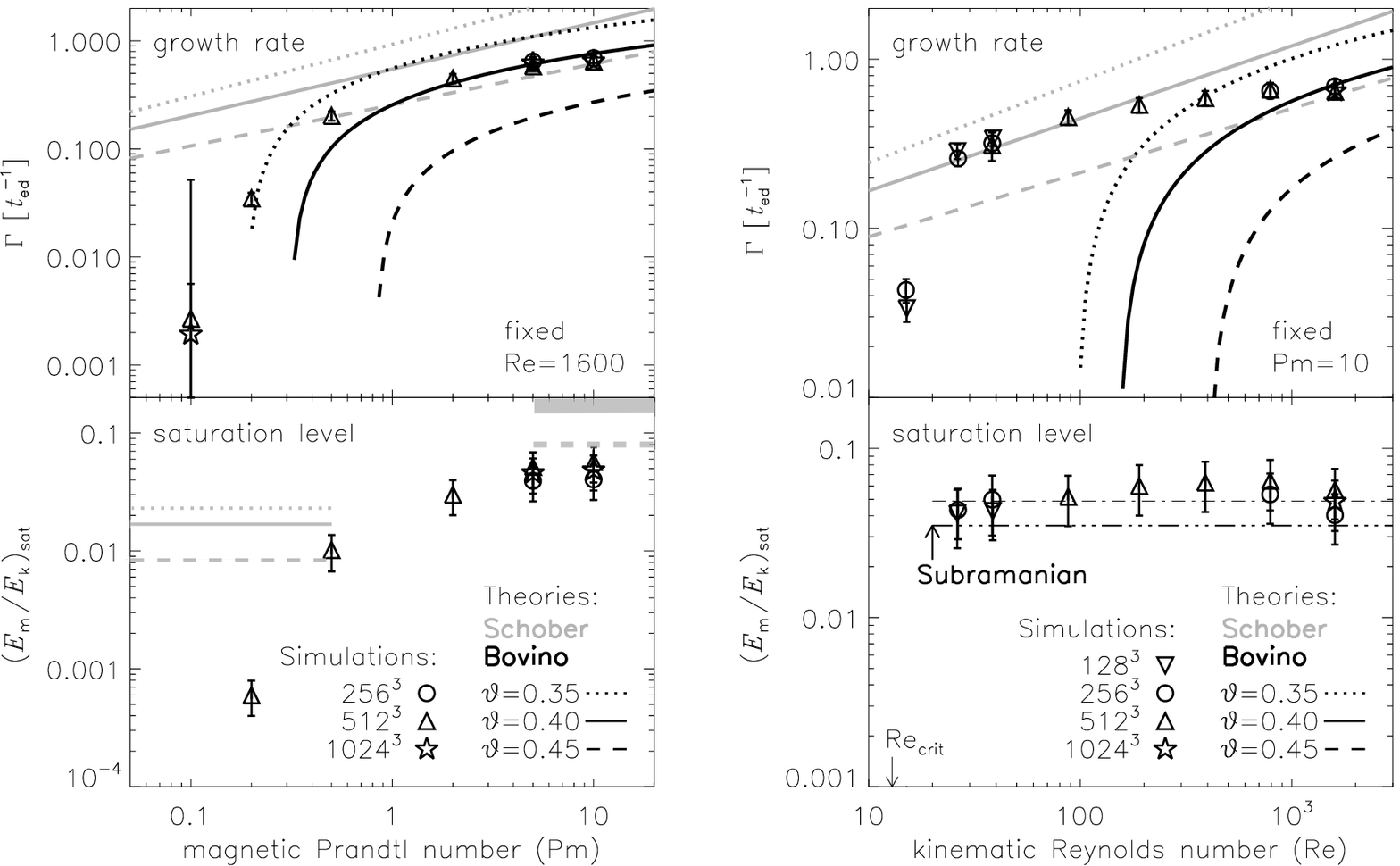}}
\caption{\emph{Left panels:} dynamo growth rate $\Gamma$ (\emph{top}) and saturation level $\esat$ (\emph{bottom}) as a function of $\pmag$ for fixed $\re=1600$. Resolution studies with $256^3$, $512^3$ and $1024^3$ grid cells demonstrate convergence, tested for the extreme cases $\pmag=0.1$ and $10$. Theoretical predictions for $\Gamma$ by \citet{SchoberEtAl2012PRE2} and \citet{BovinoEtAl2013} and for $\esat$ by \citet{SchoberEtAl2015} are plotted with different line styles for a typical range of the turbulence scaling exponent, $\vartheta=0.35$ (dotted), $0.40$ (solid) and $0.45$ (dashed). \emph{Right panels:} same as left panels, but $\Gamma$ and $\esat$ are shown as a function of $\re$ for fixed $\pmag=10$. The dot-dashed line is a fit to the simulations, yielding a constant saturation level of $\esat=0.05\pm0.01$ for $\re>\re_\mathrm{crit}\equiv\rmag_\mathrm{crit}/\pmag=12.9$ and the triple-dot-dashed line shows the result of our modified model for the saturation level originally proposed by \citet{Subramanian1999}.}
\label{fig:gratesat}
\end{figure*}

In order to compare the analytical and numerical solutions of the Kazantsev equation from \S\ref{sec:kaztheory} with the MHD simulations, we now determine the dynamo growth rate as a function of $\pmag$ for fixed $\re=1600$ and as a function of $\re$ for fixed $\pmag=10$. Depending on $\pmag$ and $\re$ we find exponential magnetic energy growth over more than 6 orders of magnitude for simulations in which the dynamo operates. We determine both the exponential growth rate $\Gamma$ and the saturation level $\esat$. The measurements are listed in table~\ref{tab:sims} and shown in figure~\ref{fig:gratesat}.

\paragraph{\bf Dependence on magnetic Prandtl number:}

In the left-hand panel of figure~\ref{fig:gratesat} we see that $\Gamma$ first increases strongly with $\pmag$ for $\pmag\lesssim1$. For $\pmag\gtrsim1$ it keeps increasing, but more slowly. The analytic and semi-analytic models by \citet{SchoberEtAl2012PRE2} and \citet{BovinoEtAl2013} both predict an increasing growth rate with $\pmag$. These models are shown for three different turbulence scaling exponents, $\vartheta=0.35$ (dotted lines), $\vartheta=0.40$ (solid lines), and $\vartheta=0.45$ (dashed lines) in equation~(\ref{eq:v_inertial}). We note that the case with a turbulence scaling exponent $\vartheta=0.45$ (the dashed lines) is the most appropriate here, because the turbulence in all our simulations is supersonic with Mach numbers in the range 4--11, for which previous high-resolution numerical simulations are consistent with $\vartheta\approx0.45$ \citep{KritsukEtAl2007,FederrathDuvalKlessenSchmidtMacLow2010}. The purely analytical solution of the Kazantsev by \citet{SchoberEtAl2012PRE2} yields power laws for $\rmag>\rmag_\mathrm{crit}$. However, the semi-analytic solution by \citet{BovinoEtAl2013} yields a sharp cutoff for $\pmag\lesssim1$, closer to the results of our 3D MHD simulations.

The theoretical predictions are qualitatively consistent with the MHD simulations. Quantitative discrepancies arise because the theoretical models assume zero helicity, $\delta$-correlation of the turbulence in time, and do not distinguish solenoidal and compressible modes in the turbulent velocity field. Finite time correlations, however, do not seem to affect the Kazantsev model significantly \citep{BhatSubramanian2014}. We therefore conclude that one needs to distinguish between the solenoidal and compressible modes in future dynamo theories, because the dynamo is primarily driven by solenoidal modes and the amount of vorticity strongly depends on the driving and Mach number of the turbulence \citep[][see \S\ref{sec:mach}]{MeeBrandenburg2006,FederrathEtAl2011PRL}.

The saturation level as a function of $\pmag$ is shown in the bottom left-hand panel of figure~\ref{fig:gratesat}. It increases with $\pmag$ similar to the growth rate and is also well converged with increasing numerical resolution. For $\pmag\gtrsim10$, $\esat$ seems to become independent of $\pmag$ (and thus independent of $\rmag$, because we have constant $\re=1600$ here). The theoretical predictions for $\esat$ from \citet{SchoberEtAl2015} are shown in the low-$\pmag$ and high-$\pmag$ limits---as for the growth rate, $\vartheta=0.45$ is the most relevant case for this comparison. The analytic prediction agrees qualitatively with the results of the MHD simulations, but similar to the limitations of the theories for the growth rate, more work is needed to incorporate the mode mixture (solenoidal vs.~compressive) in the saturation models.

\citet{Brandenburg2014} performed similar simulations and investigated the saturation level, but in the subsonic regime of turbulence (with $\mach\sim0.1$), while here we study Mach 10 MHD turbulence. \citet{Brandenburg2014} found that $\esat$ is independent of $\pmag$ for fixed $\rmag$. Here instead, we show that $\esat$ increases with $\pmag$ for fixed $\re=1600$, which implies that $\esat$ grows with $\rmag$ for $\pmag\lesssim10$. This is qualitatively consistent with the simulations (e.g., model X3 versus Y7) in \citet{Brandenburg2014}.

\paragraph{\bf Dependence on kinematic Reynolds number:}
The right-hand panels of figure~\ref{fig:gratesat} show the growth rate and saturation level as a function of kinematic Reynolds number $\re$. Similar to the dependence on $\pmag$, we find a non-linear increase in $\Gamma$ with $\re$, which is qualitatively reproduced by the semi-analytic solution of the Kazantsev equation in \citet{BovinoEtAl2013}. However, the critical Reynolds number for dynamo action is much lower in the MHD simulations than predicted by the theoretical models, which may have the same reasons as the discrepancy found for the dependence on $\pmag$, i.e., the lack of distinction of mixed turbulence modes in the Kazantsev model.

The top panels of figure~\ref{fig:gratesat} show that the growth rate ($\Gamma$) depends on both $\pmag$ and $\re$. To take both dependences into account and to determine the critical magnetic Reynolds number for dynamo action, we perform fits with the empirical model function,
\begin{equation}
\Gamma = \beta \left[ \ln(\pmag) + \ln(\re) \right] - \gamma,
\label{eq:fit_apjl}
\end{equation}
using the fit parameters $\beta$ and $\gamma$, which are related to the critical magnetic Reynolds number $\rmag_\mathrm{crit} = \exp\left(\gamma/\beta\right)$. From the fits with equation~(\ref{eq:fit_apjl}) to all our simulations we find that dynamo action is suppressed for $\rmag < \rmag_\mathrm{crit} = 129^{+43}_{-31}$ in highly compressible, supersonic MHD turbulence. Our result for $\rmag_\mathrm{crit}$ in the supersonic simulations is significantly higher than $\rmag_\mathrm{crit}$ measured in subsonic, incompressible turbulence by \citet{HaugenBrandenburgDobler2004}, who found \mbox{$\rmag_\mathrm{crit}\approx20$--$40$} for $\pmag\gtrsim1$. It is also higher than in mildly compressible turbulence, where $\rmag_\mathrm{crit}\approx50$ for $\pmag=5$ and $\mach\approx2$ \citep{HaugenBrandenburgMee2004}. The reason for the higher $\rmag_\mathrm{crit}$ compared to incompressible turbulence could be the sheet-like than vortex-like structure of supersonic turbulence \citep{Boldyrev2002,SchmidtFederrathKlessen2008,FederrathKlessenSchmidt2009} and the reduced fraction of solenoidal modes \citep[][see \S\ref{sec:mach}]{MeeBrandenburg2006,FederrathDuvalKlessenSchmidtMacLow2010,FederrathEtAl2011PRL}. The difference of the MHD simulations compared to the Kazantsev models is primarily in $\rmag_\mathrm{crit}$. \citet{BovinoEtAl2013} predicted a much higher $\rmag_\mathrm{crit}\approx4100$ for $\vartheta=0.45$. However, fits to their theoretical model yield \mbox{$\beta=0.11$--$0.19$}, which is in agreement with the range found in our MHD simulations ($\beta=0.141\pm0.004$). This demonstrates that the discrepancy between the simulations and the Kazantsev model is primarily in the critical magnetic Reynolds number, while the qualitative behaviour (determined by the $\beta$ parameter) is reproduced in the Kazantsev model.

The saturation level shown in the bottom right-hand panel of figure~\ref{fig:gratesat} is consistent with a constant level of $\esat=0.05\pm0.01$ for $\re > \re_\mathrm{crit}\equiv\rmag_\mathrm{crit}/\pmag=12.9$. Given our measurement of $\rmag_\mathrm{crit}=129$, we can compute the theoretical prediction by \citet{Subramanian1999} for the saturation level, $\esat=(3/2) (L/u_\mathrm{turb})\tau^{-1}\rmag_\mathrm{crit}^{-1}\approx0.01$. This is significantly smaller than our simulation result, assuming that $\tau=\ted=L/u_\mathrm{turb}$ is the turbulent crossing time on the largest scales of the system. However, Subramanian noted that the timescale $\tau$ is an ``unknown model parameter''. A more appropriate timescale for saturation may be the eddy-turnover timescale on the viscous scale, $\ell_\nu = L\re^{-1/(\vartheta+1)}$, for a given turbulent velocity scaling following equation~(\ref{eq:v_inertial}), because this is where the field saturates first. We find $\tau(\ell_\nu)=\ell_\nu/v(\ell_\nu)=\ted\re^{(\vartheta-1)/(\vartheta+1)}$ and with $\re=\re_\mathrm{crit}=12.9^{+4.3}_{-3.1}$, we obtain $\esat=0.035\pm0.005$ for a typical range of the velocity scaling exponent $\vartheta=0.4\pm0.1$ from molecular cloud observations \citep[e.g.,][]{Larson1981,OssenkopfMacLow2002,HeyerBrunt2004,RomanDuvalEtAl2011} and simulations of supersonic turbulence \citep{KritsukEtAl2007,SchmidtEtAl2009,FederrathDuvalKlessenSchmidtMacLow2010,Federrath2013}. The saturation level of our MHD simulations agrees within the uncertainties with our modified version of Subramanian's saturation model.

\subsubsection{Evolution of magnetic energy spectra}

\begin{figure}
\centerline{\includegraphics[width=0.60\linewidth]{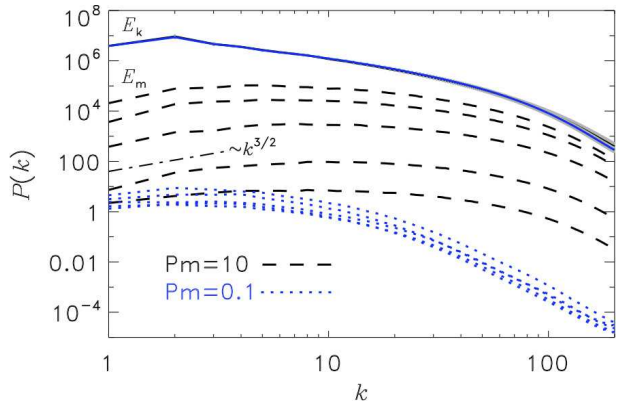}}
\caption{Time evolution of the magnetic energy power spectra for simulations with $\pmag=0.1$ (dotted lines; from bottom to top: $t/\ted=2$, $5$, $10$, $15$, $18$) and $\pmag=10$ (dashed lines; from bottom to top: $t/\ted=2$, $5$, $10$, $15$, $24$). Note that for $\pmag=10$, the last magnetic energy spectrum ($t=24\,\ted$) has just reached saturation on small scales---the $\pmag=0.1$ runs did not reach saturation within the limited computing time available, because the growth rates are extremely small for this model; cf.~figure~\ref{fig:gratesat}). The Kazantsev spectrum ($P\propto k^{3/2}$) is shown as a dash-dotted line for comparison. The solid lines show the time-averaged kinetic energy spectra.}
\label{fig:spectra}
\end{figure}

In figure~\ref{fig:spectra} we show the time evolution of the magnetic energy power spectra in our simulations with $\pmag=0.1$ and $\pmag=10$ and numerical resolution of $1024^3$ grid points. They are qualitatively consistent with incompressible dynamo studies \citep{BrandenburgSubramanian2005,MasonEtAl2011,BhatSubramanian2013}. We see that the power spectra for $\pmag=0.1$ dissipate on larger scales (lower $k$) than the $\pmag=10$ spectra, consistent with the theoretical expectation, by a factor of \mbox{$(10/0.1)^{1/(1+\vartheta)}\approx22$--$27$} for our \mbox{$\vartheta\approx0.4$--$0.5$}. Even for $\pmag=0.1$, we see the characteristic increase in the magnetic energy on all scales. The magnetic energy spectra roughly follow the Kazantsev spectrum ($\approx k^{3/2}$) on large scales \citep{Kazantsev1968,BhatSubramanian2014} in the $\pmag=10$ case, but we would expect the same to hold in the $\pmag=0.1$ case, if our simulations had larger scale separation, i.e., higher numerical resolution. The final spectrum for $\pmag=10$ has just reached saturation on small scales (approaching the kinetic energy spectrum at high $k$) and continues to grow on larger scales during the non-linear dynamo phase. The $\pmag=0.1$ runs did not have enough time to reach saturation within the limited available compute time (cf.~figure~\ref{fig:gratesat}), but we expect a non-linear dynamo phase for $\pmag<1$ that is qualitatively similar to the $\pmag>1$ case.

\subsection{Summary} \label{sec:conclusions}

In this section we presented a quantitative comparison of the turbulent dynamo in the analytical and semi-analytical Kazantsev models by \citet{Subramanian1999}, \citet{SchoberEtAl2012PRE,SchoberEtAl2012PRE2,SchoberEtAl2015} and \citet{BovinoEtAl2013} with 3D simulations of supersonic MHD turbulence. We found that the dynamo operates at low and high magnetic Prandtl numbers, but is significantly more efficient for $\pmag>1$ than for $\pmag<1$. The Kazantsev models agree qualitatively with MHD simulations, but not quantitatively. We attribute the quantitative differences to the fact that the current dynamo theories do not take into account the varying mixture of solenoidal and compressible modes in the velocity field in the case of compressible, supersonic plasmas. An extension in this direction is of high priority, and would lead to theoretical dynamo models with predictive power for the realistic high-Reynolds number regime of many astrophysical plasmas, currently inaccessible to 3D numerical simulations.


\section{Turbulent magnetic fields in the presence of ordered guide fields} \label{sec:guidefield}

In this section we present new simulations of turbulent plasmas in the presence of a guide field---an ordered magnetic field component. This is relevant to a number of astrophysical applications such as accretion discs where an ordered magnetic field component is often observed along the rotation axis of the disc at a few scale heights above and below the disc mid plane \citep{FrankEtAl2014}. Ordered guide fields are also observed towards molecular clouds in the Milky Way arms \citep{LiEtAl2014} and in the Galactic Centre cloud G0.253+0.016 \citep{PillaiEtAl2015,FederrathEtAl2016}. We note that the ordered field on a particular scale may actually be part of a turbulent field on much larger scales. However, here we are interested in understanding the amplification and evolution of the turbulent magnetic field (pressure) as a function of the guide-field strength, on the same scale.

\subsection{Simulation parameters and initial conditions} \label{sec:methods_bturb}
The simulations follow the same equations (\ref{eq:mhd1})--(\ref{eq:mhd5}) as in the dynamo studies from \S\ref{sec:mach} and \S\ref{sec:pm}, but here we add stronger ordered guide fields $B_0$ along the $z$ axis and systematically vary the strength of $B_0$. We fix the driving of the turbulence to solenoidal driving ($\zeta=1$ in equations~\ref{eq:projection} and~\ref{eq:driving_ratio}) and we keep the average energy injection rate of the turbulence constant, resulting in a roughly constant Mach number $\mach=10$ for all simulations. We use normalised, dimensionless values of all basic variables, i.e., a mean density $\rho_0=1$, sound speed $\cs=1$, and box length $L=1$. This gives a constant total mass $M=1$ and a turbulent box crossing time $\ted=L/(2\mach\cs)=0.05$. We set the kinematic viscosity to $\nu=3.33\times10^{-3}$ and the magnetic diffusivity to $\eta=1.67\times10^{-3}$, which gives a kinematic Reynolds number of $\re=L\mach\cs/(2\nu)=1500$ and magnetic Reynolds number of $\rmag=L\mach\cs/(2\eta)=3000$. Thus, the magnetic Prandtl number is fixed to $\pmag=\rmag/\re=2$. Each simulation was run with a numerical resolution of $256^3$ grid points, but we have also tested a few cases with $512^3$ grid cells and found convergence of our results.

We run a total of 14 simulations. The ordered guide field ($\Bzero$) is varied over about 5 orders of magnitude from $1.2\times10^{-2}$ to $7.1\times10^2$, which yields Alfv\'en Mach numbers with respect to the guide field in the range $\machazero=\mach\cs\sqrt{4\pi\rho_0}/B_0=0.05$--$3000$, i.e., covering the full range from sub-Alfv\'enic to super-Alfv\'enic turbulence. Since the simulation domain is periodic, the guide-field strength remains globally constant throughout the whole simulation box because of magnetic-flux conservation. The basic simulation parameters and the derived turbulent magnetic field strengths ($\Bturb$) are listed in table~\ref{tab:bturb}.

\begin{table}
\begin{center}
\begin{tabular*}{\linewidth}{@{\extracolsep{\fill} }lcccc}
Simulation Model & $\Bzero$ & $\mach$ & $\machazero$ & $\Bturb$ \\[3pt]
         \texttt{256sMS10MA3000} &           $1.2\!\times\!10^{-2}$ &   $(1.0\pm0.1)\!\times\!10^{+1}$ &   $(3.1\pm0.3)\!\times\!10^{+3}$ &   $(5.1\pm0.5)\!\times\!10^{+0}$ \\
         \texttt{256sMS10MA1000} &           $3.5\!\times\!10^{-2}$ &   $(1.0\pm0.1)\!\times\!10^{+1}$ &   $(1.0\pm0.1)\!\times\!10^{+3}$ &   $(5.2\pm0.3)\!\times\!10^{+0}$ \\
          \texttt{256sMS10MA300} &           $1.2\!\times\!10^{-1}$ &   $(1.0\pm0.1)\!\times\!10^{+1}$ &   $(3.1\pm0.3)\!\times\!10^{+2}$ &   $(4.8\pm0.7)\!\times\!10^{+0}$ \\
          \texttt{256sMS10MA100} &           $3.5\!\times\!10^{-1}$ &   $(1.0\pm0.1)\!\times\!10^{+1}$ &   $(1.0\pm0.1)\!\times\!10^{+2}$ &   $(6.5\pm0.3)\!\times\!10^{+0}$ \\
           \texttt{256sMS10MA50} &           $7.1\!\times\!10^{-1}$ &   $(1.0\pm0.1)\!\times\!10^{+1}$ &   $(5.1\pm0.5)\!\times\!10^{+1}$ &   $(7.8\pm0.4)\!\times\!10^{+0}$ \\
           \texttt{256sMS10MA20} &           $1.8\!\times\!10^{+0}$ &   $(1.0\pm0.1)\!\times\!10^{+1}$ &   $(2.0\pm0.2)\!\times\!10^{+1}$ &   $(1.1\pm0.1)\!\times\!10^{+1}$ \\
           \texttt{256sMS10MA10} &           $3.5\!\times\!10^{+0}$ &   $(9.8\pm1.0)\!\times\!10^{+0}$ &   $(9.8\pm1.0)\!\times\!10^{+0}$ &   $(1.4\pm0.1)\!\times\!10^{+1}$ \\
            \texttt{256sMS10MA5} &           $7.1\!\times\!10^{+0}$ &   $(9.3\pm0.9)\!\times\!10^{+0}$ &   $(4.7\pm0.5)\!\times\!10^{+0}$ &   $(1.9\pm0.1)\!\times\!10^{+1}$ \\
            \texttt{256sMS10MA2} &           $1.8\!\times\!10^{+1}$ &   $(9.0\pm0.9)\!\times\!10^{+0}$ &   $(1.8\pm0.2)\!\times\!10^{+0}$ &   $(2.2\pm0.1)\!\times\!10^{+1}$ \\
            \texttt{256sMS10MA1} &           $3.5\!\times\!10^{+1}$ &   $(1.0\pm0.1)\!\times\!10^{+1}$ &   $(1.0\pm0.1)\!\times\!10^{+0}$ &   $(1.8\pm0.1)\!\times\!10^{+1}$ \\
           \texttt{256sMS10MA05} &           $7.1\!\times\!10^{+1}$ &   $(1.1\pm0.1)\!\times\!10^{+1}$ &   $(5.6\pm0.6)\!\times\!10^{-1}$ &   $(1.2\pm0.1)\!\times\!10^{+1}$ \\
           \texttt{256sMS10MA02} &           $1.8\!\times\!10^{+2}$ &   $(1.5\pm0.1)\!\times\!10^{+1}$ &   $(2.9\pm0.3)\!\times\!10^{-1}$ &   $(8.5\pm1.3)\!\times\!10^{+0}$ \\
           \texttt{256sMS10MA01} &           $3.5\!\times\!10^{+2}$ &   $(1.6\pm0.2)\!\times\!10^{+1}$ &   $(1.6\pm0.2)\!\times\!10^{-1}$ &   $(5.6\pm1.2)\!\times\!10^{+0}$ \\
          \texttt{256sMS10MA005} &           $7.1\!\times\!10^{+2}$ &   $(1.7\pm0.2)\!\times\!10^{+1}$ &   $(8.4\pm0.8)\!\times\!10^{-2}$ &   $(3.3\pm0.8)\!\times\!10^{+0}$
\end{tabular*}
\caption{List of turbulence simulations with different guide-field strength ($B_0$).}
\label{tab:bturb}
\end{center}
\end{table}

\subsection{Results and discussion}

\begin{figure}
\centerline{\includegraphics[width=0.75\linewidth]{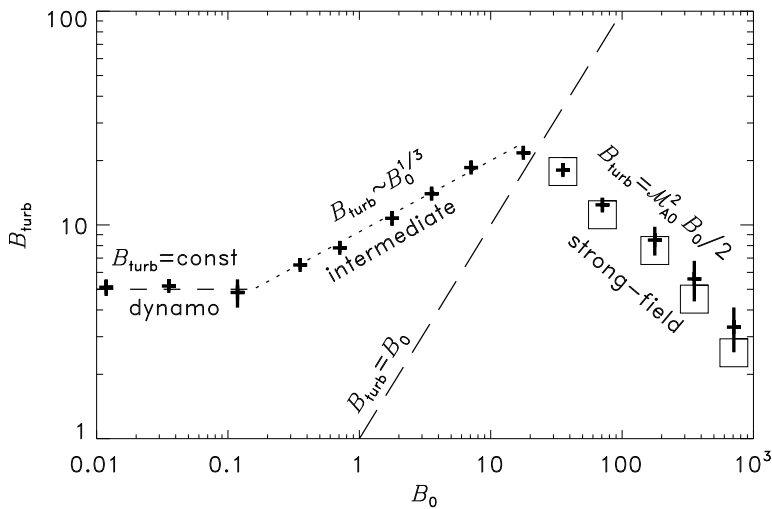}}
\caption{Turbulent (un-ordered) magnetic field ($\Bturb$) as a function of (ordered) magnetic guide field ($B_0$) from the simulations (crosses with 1$\sigma$ error bars) listed in table~\ref{tab:bturb}. We can distinguish three regimes: i) the dynamo regime for weak $B_0$, ii) the intermediate regime, and iii) the strong guide-field regime. The dashed and dotted lines indicate fits of $\Bturb$ for the dynamo regime and the intermediate regime, respectively. The boxes show the theoretical estimate of $\Bturb$ in the strong-field regime (equation~\ref{eq:sgf}). The long-dashed line shows $\Bturb=B_0$, which separates the intermediate from the strong-field regime.}
\label{fig:bturb}
\end{figure}

As for the simulation models in \S\ref{sec:mach} and \S\ref{sec:pm}, we evolve the simulations in table~\ref{tab:bturb} until they reach a converged state in the turbulent magnetic field component ($\Bturb$). Figure~\ref{fig:bturb} shows $\Bturb$ as a function of $B_0$. We find the remarkable result that although we varied $B_0$ over 5 orders of magnitude, $\Bturb$ only varies by a factor of $\sim\!10$. Our measurements of $\Bturb$ are listed in the last column of table~\ref{tab:bturb}.

We identify three different regions in figure~\ref{fig:bturb}. First, for low guide fields, we find $\Bturb=\mathrm{const}$. This is the `dynamo regime'. Second, in the `intermediate regime' we find that $\Bturb$ increases with $B_0$ as $\Bturb\propto B_0^{1/3}$. Finally, in the third regime---the `strong-field regime'---$\Bturb$ decreases with increasing $B_0$ and follows the relation $\Bturb=\machazero^2 B_0/2$ shown by the boxes. The intermediate and strong-field regimes are separated by the line $\Bturb=B_0$. We now derive the theoretical model, $\Bturb=\machazero^2 B_0/2$, for the strong guide-field regime.

\subsection{Theoretical model for the turbulent magnetic field}

We provide a simple analytical model for $\Bturb$ in the limits of weak and strong guide field, respectively ($\Bturb\gg B_0$ and $\Bturb\ll B_0$). We start by separating the magnetic field into an ordered component ($B_0$) and an un-ordered, turbulent component ($\Bturb$),
\begin{align}
B & = B_0 + \Bturb \\
\iff\; B^2 & = B_0^2 + \Bturb^2 + 2B_0\Bturb. \label{eq:bsq}
\end{align}
Note that the mean field $\langle B \rangle = B_0$ and $\langle\Bturb\rangle = 0$, while the magnetic energy density is proportional to $\langle B^2 \rangle = B_0^2 + \langle \Bturb^2 \rangle + 2\langle B_0\Bturb \rangle$. 

\subsubsection{Weak magnetic guide field}
The limit of a weak or vanishing guide field is the dynamo limit that we explored in detail in \S\ref{sec:mach} and \S\ref{sec:pm}. In the limit $\Bturb\gg B_0$, equation~(\ref{eq:bsq}) becomes $B^2\approx\Bturb^2$ and thus the turbulent magnetic energy density $e_\mathrm{m}\approx\Bturb^2/(8\pi)$. From the models of dynamo saturation (c.f.~\S\ref{sec:mach} and \S\ref{sec:pm}), we know that the magnetic energy can only reach a fraction $\eps_\mathrm{sat}=\esat$ of the turbulent kinetic energy because of the back reaction of the field via the Lorentz force. Thus, in the dynamo limit we make the ansatz,
\begin{align}
 & e_\mathrm{m} = \eps_\mathrm{sat} e_\mathrm{k} \\
\iff\; & \Bturb^2/(8\pi) = \eps_\mathrm{sat} \rho_0 u_\mathrm{turb}^2/2 \\
\iff\; & \Bturb = \left(4\pi\eps_\mathrm{sat}\rho_0 u_\mathrm{turb}^2\right)^{1/2} \\
\iff\; & \Bturb = \left(4\pi\eps_\mathrm{sat}\rho_0 \cs^2 \mach^2\right)^{1/2}.
\end{align}
To evaluate this equation for the present case ($\rho_0=\cs=1$ and $\mach=10$; see \S\ref{sec:methods_bturb}), we take the measured saturation level $\eps_\mathrm{sat} \approx 0.02$ from the middle panel of figure~\ref{fig:prl_gratesat} for solenoidal driving at Mach 10 and find $\Bturb\approx5$. This is in excellent agreement with the simulations in the dynamo limit shown in figure~\ref{fig:bturb}.

\subsubsection{Strong magnetic guide field}
In the limit of a strong guide field ($\Bturb\ll B_0$), equation~(\ref{eq:bsq}) can be approximated as $B^2\approx B_0^2 + 2B_0\Bturb$ and the un-ordered (turbulent) magnetic energy density is $e_\mathrm{m}\approx B_0\Bturb/(4\pi)$. We now make the ansatz that this energy is provided by the turbulent kinetic energy via tangling of the ordered field component,
\begin{align}
 & e_\mathrm{m} = e_\mathrm{k} \\
\iff\; & B_0\Bturb/(4\pi) = \rho_0 u_\mathrm{turb}^2/2 \\
\iff\; & \Bturb = 2\pi\rho_0 u_\mathrm{turb}^2 / B_0 \\
\iff\; & \Bturb = \machazero^2 B_0 / 2, \label{eq:sgf}
\end{align}
where we have identified the guide-field Alfv\'en Mach number $\machazero=u_\mathrm{turb}\sqrt{4\pi\rho_0}/B_0$ in the last step. Our simulations in figure~\ref{fig:bturb} are in very good agreement with this theoretical relation for the strong guide-field limit (shown as boxes in figure~\ref{fig:bturb}).


\section{Astrophysical implications of strong magnetic fields} \label{sec:implications}

We have seen that turbulent magnetic fields can be generated and amplified over a wide range of physical conditions, from subsonic to supersonic plasmas (\S\ref{sec:mach}), with high or low magnetic Prandtl number (\S\ref{sec:pm}), and in the presence of a magnetic guide field (\S\ref{sec:guidefield}). This means that magnetic fields can potentially play a significant role in various astrophysical systems and we now discuss a few examples.

\subsection{Accretion discs and protostellar jets}

Magnetic fields play a crucial role for the dynamics of accretion discs through the magneto-rotational instability (MRI) \citep{BalbusHawley1991}, which drives turbulence and angular momentum transport, thereby allowing the central star to gain more mass. The structural changes of the disc caused by the MRI may also influence the disc's potential to fragment and form planets \citep{BaiStone2014}. In the earlier phases of protostellar accretion discs, powerful jets and outflows are launched. It is well known that this type of mechanical feedback is caused by the winding-up of the magnetic field in the rotating disc \citep{BlandfordPayne1982,LyndenBell2003,PudritzEtAl2007,KrumholzEtAl2014,FrankEtAl2014}. Jets and outflows drive turbulence \citep{NakamuraLi2007,NakamuraLi2011} and reduce the star formation rate by about a factor of 2 \citep{WangEtAl2010,Federrath2015}. They further reduce the average mass of stars by a factor of 3 \citep{FederrathEtAl2014}, thus having a strong impact on the initial mass function (IMF) of stars, the origin of which remains one of the biggest open questions in astrophysics \citep{OffnerEtAl2014}.

\subsection{The interstellar medium of galaxies and molecular cloud formation}

The importance of magnetic fields for the formation of molecular clouds has been investigated in several works \citep{PassotVazquezPouquet1995,HennebelleEtAl2008,BanerjeeEtAl2009,HeitschStoneHartmann2009,VazquezSemadeniEtAl2011,SeifriedEtAl2011}. Recently, \citet{KoertgenBanerjee2015} find suppression of star formation by moderate magnetic fields of the order of $3\,\mu\Gauss$, emphasising the role that the magnetic field plays in regulating the formation of stars.

A central outstanding problem is to explain and understand the observed relation of magnetic field strength with cloud density \citep{Crutcher2012}. The observations suggest that the magnetic field strength is fairly constant with values \mbox{$\sim1$--$10\,\mu\Gauss$} for number densities $n\lesssim100\,\cm^{-3}$. For densities greater than this, the field increases roughly as $B\propto n^{2/3}$, consistent with simulations \citep[e.g.,][]{PadoanNordlund1999,LiMcKeeKlein2015}. The problem, however, is that in order for clouds to be able to form stars, they need to cross from the so-called 'sub-critical' regime (where the collapse of gas is entirely suppressed by the magnetic field) into the 'super-critical' regime. How exactly this transition occurs is not well understood. Probably some form of magnetic flux loss is required, because otherwise, clouds that start off sub-critical at low densities (which indeed seems to be the magnetic state of most of the diffuse gas in the galaxy) would stay sub-critical forever and would not form stars.

\subsection{Star formation}

It was long thought that magnetic fields control the formation of stars in the interstellar medium through ``ambipolar diffusion'', the slow drift of neutral gas through the ionised gas towards the centre of the clouds, where star formation would occur \citep{ShuAdamsLizano1987}. In the last decade, this picture has been replaced by a turbulence-regulated theory of star formation  \citep{MacLowKlessen2004,ElmegreenScalo2004,ScaloElmegreen2004,McKeeOstriker2007,HennebelleFalgarone2012,PadoanEtAl2014}, where magnetic fields were considered somewhat less important. However, the most recent years have seen a revival of the role of magnetic fields for star formation and for the structure of the interstellar medium. Both supercomputer simulations and observations have contributed to this recent development.

Firstly, observations show that the magnetic energy density is comparable to the turbulent kinetic energy density in the interstellar medium and in some star-forming clouds \citep{StahlerPalla2004}. This is reflected in turbulent Alfv\'en Mach numbers around unity \citep{HeyerBrunt2012} and strong ordered magnetic fields on cloud scales \citep{LiHenning2011,LiEtAl2011,LiEtAl2014}, in the galactic centre \citep{DotsonEtAl2010,PillaiEtAl2015,FederrathEtAl2016} and on galactic scales \citep{BeckEtAl1996,Beck2016}.

Secondly, simulations have demonstrated that magnetic pressure reduces fragmentation, such that only about half as many stars form in the presence of a typical magnetic field compared to the case without magnetic fields \citep{FederrathKlessen2012}. This means that magnetic fields cannot be ignored for the IMF \citep{PriceBate2007,HennebelleTeyssier2008,BuerzleEtAl2011,PetersEtAl2011,HennebelleEtAl2011}. Moreover, magnetic pressure slows the formation of stars by about a factor of 2 \citep{PadoanNordlund2011,FederrathKlessen2012,Federrath2015}. Recent work suggests that strong magnetic guide fields aligned with the gas flow can also have the opposite effect and increase the star formation rate \citep{ZamoraAvilesEtAl2016}. Current star formation theories only take the un-ordered turbulent magnetic field component into account \citep{FederrathKlessen2012}. Thus, adding to these theories the effects of magnetic tension caused by strong ordered fields is a priority for the near future.



\vspace{0.6cm}
The author thanks the two anonymous referees for their comments, which helped to improve this work.
The author gratefully acknowledges funding provided by the Australian Research Council's Discovery Projects (grant~DP150104329).
The simulations presented in this work used high performance computing resources provided by the Leibniz Rechenzentrum and the Gauss Centre for Supercomputing (grants~pr32lo, pr48pi and GCS Large-scale project~10391), the Partnership for Advanced Computing in Europe (PRACE grant pr89mu), the Australian National Computational Infrastructure (grant~ek9), and the Pawsey Supercomputing Centre with funding from the Australian Government and the Government of Western Australia, in the framework of the National Computational Merit Allocation Scheme and the ANU Allocation Scheme.
The simulation software FLASH was in part developed by the DOE-supported Flash Center for Computational Science at the University of Chicago.



\end{document}